\documentclass[fleqn,usenatbib]{mnras}

\usepackage{graphicx}	
\usepackage{amsmath}	
\usepackage{amssymb}	
\usepackage{multicol}        
\usepackage{bm}		
\usepackage{natbib}
\usepackage{threeparttable}
\usepackage{aas_macros}
\def\gizmo{\texttt{GIZMO}}
\def\ahf{\texttt{AHF}}

\def\fire{\texttt{FIRE}}
\defcitealias{Hopkins:2017}{H17}
\newcommand{\kms}{{\rm km\,s^{-1}}}
\newcommand{\rh}{\mathrm{r_{1/2}}}

\newcommand{\lcdm}{$\Lambda$CDM}
\newcommand{\msun}{{\rm M}_{\odot}}
\newcommand{\mhalo}{M_{\rm halo}}
\newcommand{\mdyn}{M_{\rm dyn}}
\newcommand{\mvir}{M_{\rm vir}}
\newcommand{\rvir}{R_{\rm vir}}
\newcommand{\mstar}{M_{\star}}
\newcommand{\vmax}{V_{\rm max}}
\newcommand{\vcirc}{V_{\rm circ}}
\newcommand{\mpc}{{\rm Mpc}}
\newcommand{\hopkins}{\citetalias{Hopkins:2017}}
\newcommand{\rpower}{r_{\rm power}}

\usepackage[T1]{fontenc}
\usepackage{ae,aecompl}

\usepackage{newtxtext,newtxmath}

\title[\texttt{FIRE} in the Field]{\texttt{FIRE} in the Field: Simulating the Threshold of Galaxy Formation}

\author[A. Fitts et al.]{Alex Fitts$^1$\thanks{\href{mailto:afitts@astro.as.utexas.edu}{afitts@astro.as.utexas.edu}}, 
Michael Boylan-Kolchin$^1$\thanks{\href{mailto:mbk@astro.as.utexas.edu}{mbk@astro.as.utexas.edu}}, Oliver D. Elbert$^2$, James S. Bullock$^2$, \newauthor
Philip F. Hopkins$^3$, Jose O\~norbe$^4$, Andrew Wetzel$^{3, 5, 6}\thanks{Caltech-Carnegie Fellow}$, Coral Wheeler$^{2,3}$, \newauthor
Claude-Andr\'e Faucher-Gigu\`ere$^7$, Du\v{s}an Kere\v{s}$^8$, Evan D. Skillman$^{9}$, Daniel R. Weisz$^{10}$
\\
$^1$The University of Texas at Austin, Department of Astronomy, 2515 Speedway, Stop C1400, Austin, Texas 78712-1205\\
$^2$Department of Physics and Astronomy, Center for Cosmology, 4129 Reines Hall, University of California Irvine, CA 92697, USA\\
$^3$TAPIR, California Institute of Technology, Pasadena, CA, USA\\
$^4$Max-Planck-Institut f\"ur Astronomie, K\"onigstuhl 17, 69117 Heidelberg, Germany\\
$^5$Carnegie Observatories, Pasadena, CA, USA\\
$^6$Department of Physics, University of California, Davis, CA, USA\\
$^7$Department of Physics and Astronomy and CIERA, Northwestern University, Evanston, IL, USA\\
$^8$Department of Physics, Center for Astrophysics and Space Sciences, University of California, San Diego, La Jolla, CA, USA\\
$^9$Minnesota Institute for Astrophysics, University of Minnesota, 116 Church Street, SE Minneapolis, MN, 55455 USA\\
$^{10}$Department of Astronomy and Theoretical Astrophysics Center, University of California, Berkeley, CA, USA
}
\date{\today}

\pubyear{2017}

\begin{document}

\label{firstpage}
\pagerange{\pageref{firstpage}--\pageref{lastpage}}
\maketitle

\begin{abstract}
We present a suite of 15 cosmological zoom-in simulations of isolated  dark matter halos, all with masses of $M_{\rm halo} \approx 10^{10}\,\msun$ at $z=0$, in order to understand the relationship between halo assembly, galaxy formation, and feedback's effects on the central density structure in dwarf galaxies. These simulations are part of the Feedback in Realistic Environments (\fire) project and are performed at extremely high resolution ($m_{\rm baryon}=500\,\msun$, $m_{\rm dm} = 2500\,\msun$). The resultant galaxies have stellar masses that are consistent with rough abundance matching estimates, coinciding with the faintest galaxies that can be seen beyond the virial radius of the Milky Way ($\mstar/\msun \approx 10^5-10^7$). This non-negligible spread in stellar mass at $z=0$ in halos within a narrow range of virial masses is strongly correlated with central halo density or maximum circular velocity $\vmax$, both of which are tightly linked to halo formation time. Much of this dependence of $\mstar$ on a second parameter (beyond $M_{\rm halo}$) is a direct consequence of the $M_{\rm halo} \sim 10^{10}\,\msun$  mass scale coinciding with the threshold for strong reionization suppression: the densest, earliest-forming halos remain above the UV-suppression scale throughout their histories while late-forming systems fall below the UV-suppression scale over longer periods and form fewer stars as a result. In fact, the latest-forming, lowest-concentration halo in our suite fails to form \textit{any} stars. Halos that form galaxies with $\mstar \ga 2\times 10^{6}\,\msun$ have reduced central densities relative to dark-matter-only simulations, and the radial extent of the density modifications is well-approximated by the galaxy half-mass radius $\rh$. Lower-mass galaxies do not modify their host dark matter halos at the mass scale studied here. This apparent stellar mass threshold of $\mstar \approx 2\times 10^{6} \approx 2\times 10^{-4} \,\mhalo$ is broadly consistent with previous work and provides a testable prediction of FIRE feedback models in \lcdm.  
\end{abstract}



\section{Introduction}
There is strong evidence in support of the now-standard dark energy + dark matter (\lcdm) cosmological model -- or a model that reproduces \lcdm\ phenomenology -- on large cosmological scales (linear scales larger than $\sim 1\,\mpc$). On smaller scales, tests are substantially more difficult and less conclusive. The difficulty is two-fold: these small scales are firmly in the non-linear regime of cosmological density perturbations at $z=0$, meaning analytic approaches that are appropriate and straightforward for large scales no longer apply, and the galaxies that trace non-linear structure on small scales (dwarf galaxies) are inherently low-luminosity and small, making them difficult to study over cosmological scales.

Over the past two decades, improvements in instrumentation and observations have provided dramatically improved data on the internal dynamics and stellar and gaseous content of dwarf galaxies. At the same time, numerical simulations of cosmological structure formation and galaxy evolution have enabled theoretical predictions related to the abundance and structure of dwarfs. This combined progress has sharpened our view of small-scale cosmological structure and brought to light several potential discrepancies between theoretical predictions and observations: dwarf galaxies are generally less dense and less abundant than might be naively expected in \lcdm. This is the origin of the well-known Cusp/Core \citep{Moore:1994,Flores:1994}, Missing Satellites (\citealt{Klypin:1999,Moore:1999,Klypin:2015}; see also \citealt{Kauffmann:1993}), and Too Big to Fail \citep{Boylan-Kolchin:2011,Papastergis:2015a} problems. 

Much of this disagreement comes from comparing dissipationless \lcdm\ simulations with observations, with the justification (implicit or explicit) being that observations point to an increasing dominance of dark over luminous matter for increasingly faint galaxies (e.g., \citealt{McConnachie:2012}). More recent efforts to model baryonic physics in simulations of dwarf galaxies have cast significant doubt on this justification, however: many groups now find that star formation feedback can significantly affect the density structure of low-mass galaxies even if the gravitational potential is dark-matter-dominated (\citealt{Zolotov:2012,Pontzen:2012,Madau:2014,Chan:2015,Read:2016a,Tollet:2016}; see \citealt{Navarro:1996a} for earlier work on this topic). Coupled with feedback from the cosmic UV background, star formation feedback also limits the fuel for star formation, partially explaining inefficient galaxy formation in low-mass systems (and their low cosmic abundance compared to the dark matter halo mass function). 

Nevertheless, many questions remain regarding our understanding of dwarf galaxy formation, the connection between dwarf galaxies and their dark matter halos, and the use of these low-mass systems as cosmological probes. While multiple groups are now able to reproduce many properties of dwarf galaxies in numerical simulations, the input physics and other predicted properties are mutually inconsistent. For example, \citet{Zolotov:2012}, the APOSTLE simulations \citep{Sawala:2016}, and the Latte simulation \citep{Wetzel:2016} all find agreement with various observed properties of Local Group satellite galaxies, yet APOSTLE galaxies do not form cores while galaxies in \citet{Zolotov:2012} and \citet{Wetzel:2016} do. The APOSTLE simulations assume a metallicity-dependent gas density threshold for star formation of $n_{\rm sf}=0.1\,{\rm cm^{-3}}\,(Z/0.002)^{-0.64}$, with an upper limit of $n_{\rm sf}^{\rm max}=10\,{\rm cm^{-3}}$ (the same model as is used in the large volume, lower-resolution {\tt EAGLE} simulation project of \citealt{Schaye:2015}), while \citet{Zolotov:2012} use $n_{\rm sf}=100\,{\rm cm^{-3}}$ and \citet{Wetzel:2016} adopt $n_{\rm sf}=1000\,{\rm cm^{-3}}$. The implementations of star formation feedback also vary substantially across these simulations. 

While these differences may appear to be mundane and limited to details of the simulations, the stakes are actually quite high: dwarf galaxies provide critical tests of the nature of dark matter, but it is clear that we must understand the coupling between galaxy formation and dark matter dynamics if we are to test \lcdm. For example, work starting with \citet{Pontzen:2012} has shown that high star formation density thresholds (comparable to those observed in molecular clouds) are crucial for producing the bursty star formation that drives rapid gravitational potential fluctuations, which are seemingly required for core formation in \lcdm\ simulations. The formation of cores or preservation of cusps in the different simulations highlights where \lcdm+baryon predictions diverge and the necessity of modeling star formation in the most realistic manner possible. This is particularly true for dwarf galaxies, which have long been known to be sensitive to supernova feedback and the effects of cosmic reionization \citep[e.g.,][]{Dekel:1986,Efstathiou:1992,Babul:1992,Bullock:2000,Somerville:2002,Benson:2002}. Additional sources of heating such as cosmic rays \citep{Chen:2016} and TeV blazars \citep{Pfrommer:2012} may also be important but are not often modeled in numerical simulations.

Current results point to $\mhalo(z=0) \sim 10^{10}\,\msun$ as a crucial mass scale for understanding dwarf galaxies and their consistency with \lcdm. It is the characteristic mass scale at $z=0$ at which the baryon fraction of halos is reduced by 50\% relative to the cosmic baryon fraction $f_{\rm b} \equiv \Omega_{\rm b}/\Omega_{\rm m}$ (which is $0.168$ for the cosmology we adopt in this paper, as detailed at the end of this section) owing to the cosmic UV background \citep{Hoeft:2006,Okamoto:2008,Noh:2014}. Halos at this mass scale are therefore likely to serve as sensitive probes of reionization-induced feedback, which may contribute to the diversity of star formation histories observed for low-mass galaxies \citep[e.g.,][]{Tolstoy:2009,Skillman:2014,Weisz:2014a,Brown:2014,Gallart:2015,Skillman:2017}. Counts of galaxies in the Local Group, coupled with numerical simulations, also point to this as the crucial halo mass (or an equivalent peak circular velocity of $40\,\kms$) at which stellar feedback switches from being efficient (at higher $\mhalo$) to inefficient (at lower $\mhalo$) at redistributing dark matter in galaxies' centers \citep{Governato:2012,DiCintio:2014,Onorbe:2015,Tollet:2016,Kormendy:2016}. 

Based on simulations and extrapolated $\mstar-\mhalo$ relations, the stellar  content of these halos is expected to be $\mstar \sim 10^6\,\msun$ \citep{Munshi:2013,Ferrero:2012,Moster:2013,Behroozi:2013b,Tollet:2016}, comparable to the stellar masses of classical dwarf spheroidal satellites in the Local Group that can be studied in exquisite detail observationally. These objects, and this stellar mass scale, are directly related to the cusp/core and too big to fail problems, as well to understanding the baryonic Tully-Fisher relation and deviations from it on  dwarf scales \citep{McGaugh:2010,Sales:2016}. 
The $\mhalo \sim 10^{10}\,\msun$ mass scale is therefore a unique probe of connections among star formation and feedback, cosmic reionization, and dark matter physics.

Most modern high-resolution simulation suites are designed to cover a wide range of halo masses \citep{Munshi:2013,Chan:2015,Wang:2015,Sawala:2016}. This broad strategy is clearly essential for a general understanding of galaxy formation, but it is not well-matched to understanding this crucial halo mass scale. In this paper, we eschew a broad approach to focus on galaxies that form in halos with $\mhalo(z=0)\approx10^{10}\,\msun$ through cosmological zoom-in simulations \citep{Katz:1993,Onorbe:2014}. Our halos are selected to be isolated (non-satellites) and span a range of assembly histories, allowing us to test connections among halo assembly, galaxy formation, feedback, and the central dark matter content of dwarf galaxies. All of our simulations use the \gizmo\ code \citep{Hopkins:2015} and the FIRE-2 model for galaxy formation and feedback (Hopkins et al. \citeyear{Hopkins:2014} and \citeyear{Hopkins:2017}); further details, along with an overview of our simulation suite, are given in  Section~\ref{sec:simulations}. Section~\ref{sec:results} presents our primary results, including the dependence of stellar content and potential dark matter core formation on dark matter halo assembly. Section~\ref{sec:discussion} provides a synthesis of our results in the context of current understanding of dwarf galaxy formation and evolution. Section~\ref{sec:summary} gives a summary of our main findings. Our work is based on the \lcdm\ model with parameters taken from analysis of the \textit{Wilkinson Microwave Anisotropy Probe} 7-year data \citep{Komatsu:2011}: $h = 0.71$, $\Omega_{\rm m} = 0.266$, $\Omega_{\rm b} = 0.0449$, $\Omega_\Lambda = 0.734$, $n_{\rm s}$ = 0.963, and $\sigma_8$ = 0.801.

\section{Simulations}
\label{sec:simulations}
Our simulation suite consists of 15 zoom-in simulations of \lcdm\ dark matter halos chosen to have virial\footnote{We define all virial quantities using the \citet{Bryan:1998} definition of the virial overdensity. For our chosen cosmology, $\Delta_{\rm vir}=96.45$ (relative to $\rho_{\rm crit}$) and $\mvir=10^{10}\,M_\odot$ corresponds to $R_\mathrm{vir}\approx56$ kpc at $z=0$.} masses of $10^{10}\,\msun \:(\pm 30\%)$ at $z=0$. 12 of the 15 halos were selected from parent simulations of homogeneously-resolved volumes with side lengths of $25 \,h^{-1}\,\mpc$. The other three halos were selected from parent volumes with side lengths of $5 \,h^{-1}\,\mpc$. 
To ensure that we explore the physics of star formation and internal feedback separately from environmental effects, each target halo is required to be separated from any more massive halo by at least 3 times the virial radius of the more massive halo (while any more massive halo is required to lie beyond 5 times the virial radius of the target halo). The halos span a representative range of concentrations (and therefore, formation times; e.g., \citealt{Navarro:1997,Wechsler:2002}) for their mass. Initial conditions are generated with \texttt{MUSIC} \citep{Hahn:2011}. 

All of our simulations are run using the \gizmo\footnote{\url{http://www.tapir.caltech.edu/~phopkins/Site/GIZMO.html}} code \citep{Hopkins:2015}. Our fiducial simulations with galaxy formation physics included have baryonic (dark matter) particle masses of $500\,M_{\sun}$ ($2500\,M_{\sun}$), with physical baryonic (dark matter) force resolution of $h_{b}=2\,$pc ($\epsilon_{\rm DM}=35\,$pc); force softening for baryons uses the fully-conservative adaptive algorithm from \citet{Price:2007}, meaning that the gravitational force assumes the identical mass distribution as the hydrodynamic equations (resulting in identical hydrodynamic and gravitational resolution). Particle masses are a factor of 2 smaller for the three halos selected from smaller parent volumes (halos m10g, m10q, m10v). For each halo, we also simulate a dark-matter-only (DMO) version. These simulations have identical initial conditions, except the baryonic particles are subsumed into dark matter particles for the DMO run, making the individual particle masses larger by a factor of $(1-f_{\rm b})^{-1}$ (where $f_{\rm b} \equiv \Omega_{\rm b}/\Omega_{\rm m}$ is the cosmic baryon fraction). We therefore opt to quote results for the DMO simulations using $m_{\rm p} \rightarrow (1-f_{\rm b})\,m_{\rm p}$. This means we adjust $\rho(r)\rightarrow (1-f_{\rm b})\,\rho(r)$ and $V_{\rm circ}(r) \rightarrow \sqrt{1-f_{\rm b}}\,V_{\rm circ}(r)$ for all results quoted for DMO simulations unless otherwise noted, effectively mimicking maximal baryonic mass loss. For convergence-testing, we run a subset of simulations denoted "Z12" (our fiducial runs are "Z13") at $2\times$ poorer force and $8\times$ poorer mass resolution. To understand numerical convergence properly, we also simulate DMO versions for three of the halos at one level higher in resolution (Z14, $2\times$ better force and $8\times$ better mass resolution than Z13). The demands of numerical convergence, and implications of using under-resolved simulations, are discussed in Appendix~\ref{sec:appendixa}.

These simulations are part of the Feedback In Realistic Environments (\fire)\footnote{\url{http://fire.northwestern.edu}} project \citep{Hopkins:2014}. \fire~cosmological simulations of dwarf galaxies have reproduced several key observables, including realistic galactic outflows \citep{Muratov:2015}, the mass-metallicity relation \citep{Ma:2016}, the mass-size relation and age/metallicity gradients \citep{El-Badry:2016}, cored dark-matter profiles \citep{Onorbe:2015,Chan:2015}, and stellar kinematics \citep{Wheeler:2017}. Those previous papers all used the identical, original version of the \fire~code (henceforth "\fire-1"). In this work, we take advantage of recent improvements to the \fire\ code, which collectively constitute the \fire-2 model
(\citealt{Hopkins:2017}, hereafter \hopkins). The most significant change is the hydrodynamics methodology: while \fire-1 used the older pressure-energy smoothed-particle hydrodynamics ("\texttt{P-SPH}"; \citealt{Hopkins:2013}) method, \fire-2 uses the new mesh-free finite-mass (MFM) Lagrangian method in \gizmo. MFM is a second-order accurate method that maintains advantages of SPH such as excellent conservation of mass, energy, momentum, and angular momentum while also capturing advantages of grid-based methods,  including sharp shock-capturing, minimal numerical viscosity, higher-order convergence, and accurate treatment of fluid mixing. We stress that the set of physics simulated, and feedback inputs from stellar evolution models, are the same in FIRE-1 and FIRE-2. As with \fire-1, all \fire-2 simulations (e.g., \citealt{Wetzel:2016,Su:2016}) use the identical physics, source code, and numerical parameters. Extensive details of the method and numerical tests are presented in \hopkins; we therefore only briefly summarize them here.

Gas follows an ionized+atomic+molecular cooling curve from 10 to $10^{10}$ K, including metallicity-dependent fine-structure and molecular cooling at low temperatures, and high-temperature ($>10^4$ K) metal-line cooling followed species-by-species for 11 separately tracked species, with a redshift-dependent, spatially uniform UV background\footnote{We use the December 2011 update of the \citet{Faucher-Giguere:2009} UVB model, available at \url{http://galaxies.northwestern.edu/uvb}. This model is compatible with \citet{Planck:2016a} cosmological constraints, with hydrogen reionization completing by $z = 10$ and helium II reionization completing by $z \sim 3.3$.}
and local sources. At all times, we tabulate relevant ionization states and cooling rates from a compilation of \texttt{CLOUDY} runs \citep{Ferland:1998}, accounting for gas self-shielding. Star formation occurs only in locally self-gravitating (following \citealt{Hopkins:2013}), self-shielding and molecular (following \citealt{Krumholz:2011}), Jeans-unstable regions with densities $>1000\,{\rm cm^{-3}}$; gas that meets these criteria is turned into stars on its free-fall time. Star particles are taken to be simple stellar populations (known age and metallicity) with a \citet{Kroupa:2001} initial mass function. For each star particle, the simulations explicitly follow stellar feedback in the form of: (i) local and long-range momentum flux from radiation pressure (in the initial UV/optical single-scattering, and re-radiated light in the infrared); (ii) energy, momentum, mass, and metal injection from supernovae (types Ia and II) and stellar mass loss (both OB and AGB), and (iii) photo-ionization and photo-electric heating. All feedback event rates, luminosities and energies, mass-loss rates, and all other quantities are tabulated directly from stellar evolution models (STARBURST99 ver7.0; \citealt{Leitherer:1999}).

In post-processing, we identify halos and construct merger trees with the Amiga Halo Finder (\ahf; \citealt{Knollmann:2009}). We have found that the centers from our simulation outputs identified by \ahf\ can differ by as much as 200-400 pc from the centers identified by other methods. This can have serious consequences for interpretation of the central densities of galaxies and dark matter halos, as it will result in an apparent core in an inherently cuspy profile. Since the mis-centering we find is comparable to the sizes of many dwarf galaxies, it is a potentially serious issue. We therefore adopt an iterative "shrinking spheres" \citep{Klypin:1997,Power:2003,Navarro:2004} centering routine based on the \ahf\ halo catalogs that utilizes both the dark matter and star particles' positions (with weighting according to the particle mass) to recompute halo (and galaxy) centers. All profiles are constructed from these centers; centering on dark matter or stars alone gives indistinguishable results.

We fit each halo to an Einasto (\citeyear{Einasto:1965}) density profile:
\begin{equation}
\rho(x \equiv r/r_2)=\rho_2\,\exp\left[-\frac{2}{\alpha}\left(x^\alpha-1 \right) \right]\,,
\end{equation}
where $\rho_{2}$ and $r_{2}$ are the density and radius where $d\log{\rho}/d\log{r}=-2$ and $\alpha$ is a shape parameter. This is similar to the familiar \citet*[][hereafter NFW]{Navarro:1996b} density profile but provides a slightly better fit to both individual and stacked density profiles from simulations \citep{Navarro:2004,Merritt:2006,Prada:2006,Gao:2008}, even after fixing $\alpha=0.17$ (so that both models have two free parameters). We use these fits (with $\alpha$ fixed to 0.17) to calculate a concentration parameter $c_{\rm vir,\,DMO}\equiv R_{\rm vir}/r_{2}$ for each simulation. For an NFW profile, $r_2$ is equivalent to the scale radius $r_{\rm s}$; our concentration measure is therefore the familiar concentration parameter for halos well-fitted by NFW profiles.

\section{Results}
\label{sec:results}
\begin{figure}
\includegraphics[width=\columnwidth]{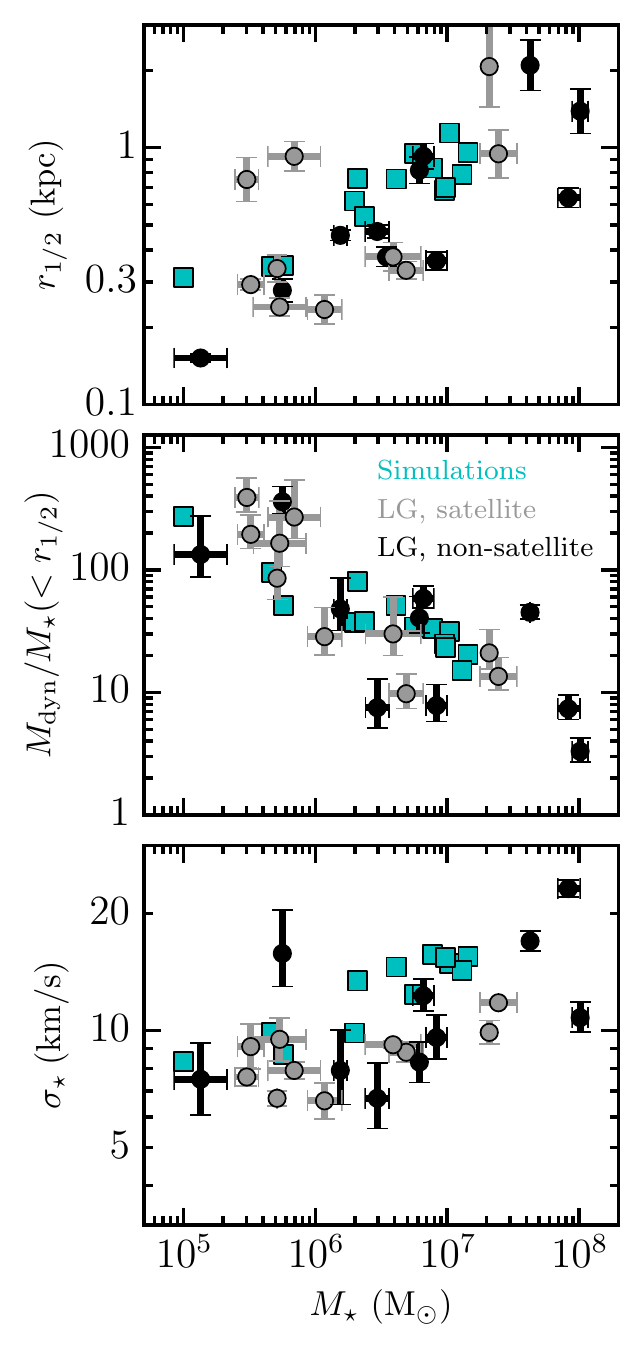}
 \caption{\textit{Top}: the 3D stellar half-mass radius $r_{1/2}$ as a function of $z=0$ stellar mass. \textit{Middle}: ratio of total (dynamical) mass to $\mstar$ within $r_{1/2}$ as a function of $\mstar(z=0)$. \textit{Bottom}: 1D stellar velocity dispersion (computed as $\sigma_{\star,\rm 3D}/\sqrt{3}$) as a function of $\mstar(z=0)$. Our simulated galaxies are plotted as cyan squares; data for observed satellite dwarf spheroidals (gray circles) and non-satellite dwarf irregular galaxies (black circles) in the Local Group (from \citealt{Kirby:2013} and \citealt{Kirby:2014}) are also plotted for comparison. In each panel, the simulations follow the same trends as the observations and fall in the same part of parameter space. }
\label{fig:correlations}
\end{figure}

\begin{table*}
  \caption{\textit{Global properties at $z=0$ for simulated field galaxies with $\mvir \approx 10^{10} \,\msun$}. Columns:
(1) Virial mass; 
(2) Maximum amplitude of rotation curve; 
(3) Stellar mass of the central galaxy [defined as $\mstar(<0.1\,\rvir)$];
(4) Mass of gas below $T=10^4$ K within $\rvir$; 
(5) Total baryon fraction within $\rvir$ scaled to cosmic baryon fraction $f_{\rm b}$; 
(6) 3D stellar half-mass radius;
(7) Ratio of total mass to stellar mass within the stellar half-mass radius; 
(8) Ratio of virial mass in hydro run to virial mass in DMO run (after correcting the DMO virial mass for $f_{\rm b}$); 
(9) Maximum of the rotation curve (DMO, after correction for $f_{\rm b}$); 
(10) Einasto concentration parameter (DMO).
\label{table:params}
}
 \begin{threeparttable}
\centering 
\begin{tabular}{lcccccccccc}
\hline
\hline  
& $M_{\mathrm{vir}}$ &$\vmax$& $\mstar$ & $\mathrm{M_{gas,cold}}$ &$f_\mathrm{baryon}/f_\mathrm{b}$& $\rh$&$\mathrm{M_{dyn}}/\mstar$ & $M_\mathrm{hydro}/M_\mathrm{dmo}$&$\vmax^{\rm DMO}$&$c_{\rm vir,DMO} $\\ 
  &  $[\msun]$ &[$\kms$] & $[\msun]$ & $[\msun]$ & --&[pc]&$(<\rh)$ & --&[$\kms$]& --\\
\hline 
Halo & (1) & (2) & (3) & (4) & (5) & (6) & (7) & (8) & (9) & (10)\\
\hline 
m10a  &$7.53\times10^9$& 30.95 & 0 & 0 &0.048 & --- & ---& 0.970& 30.13& 5.89  \\
m10v\tnote{a}  &$8.16\times10^9$& 30.45 & $1.00\times10^5$ & $ 7.49\times10^{6}$& 0.109 & $310$& 273.36  &0.929& 33.31& 10.40  \\
m10b  &$9.29\times10^9$& 31.51 & $4.65\times10^5$ & $ 6.63\times10^{6}$& 0.113 & $340$& 96.56  & 0.962& 34.75 & 15.34 \\
m10c &$8.92\times10^9$& 31.40 & $5.75\times10^5$ & $ 4.90\times10^{6}$& 0.112& $350$& 51.57 & 0.974& 35.53& 12.93  \\
m10d   &$8.43\times10^9$& 32.09 & $1.53\times10^6$ & $ 0$& 0.062 & $530$ & 68.47 & 0.975& 37.55& 18.42  \\
m10e &$\;1.02\times10^{10}$& 31.44& $1.98\times10^6$ & $ 2.16\times10^{7}$& 0.132 & $620$ & 37.53 & 0.979& 35.31 & 13.49  \\
m10q\tnote{a} &$7.82\times10^9$& 32.95& $2.08\times10^6$ & $4.49\times10^6$ & 0.062  & $760$& 91.33  & 0.963& 37.68& 18.30   \\
m10f   &$8.56\times10^9$& 35.66&$4.11\times10^6$ & $ 3.47\times10^{6}$& 0.081 & $750$& 54.14 & 0.944& 41.21& 21.84   \\
m10g\tnote{b} &$7.92\times10^9$& 32.10& $5.70\times10^6$ & $ 8.18\times10^{6}$& 0.076 & $950$ & 34.49 & 1.038& 37.34 &18.31  \\
m10h   &$\;1.28\times10^{10}$ & 37.98 & $7.80\times10^6$ & $ 1.59\times10^{7}$& 0.122 & $830$& 34.44  & 1.028& 44.22&$19.36$ \\
m10i  &$\;1.06\times10^{10}$ & 40.33& $8.01\times10^6$ & $ 0$ & 0.031 & $570$& 20.63 & 0.887& 45.99 & 23.85  \\
m10j  &$\;1.10\times10^{10}$ & 37.98&$9.74\times10^6$ & $1.07\times10^{7}$& 0.097 & $700$ & 23.51 & 0.975& 44.24&  24.01    \\
m10k  &$\;1.15\times10^{10}$ & 38.22 & $1.04\times10^7$ & $ 1.33\times10^{7}$& 0.091 &$1140$ &  32.52  & 0.960& 43.52 & 18.35 \\
m10l   &$\;1.06\times10^{10}$ & 37.62 & $1.30\times10^7$ & $ 8.22\times10^{6}$ & 0.096 & $780$& 15.40 & 0.958& 43.59& 21.94  \\
m10m   &$\;1.15\times10^{10}$ & 38.51 & $1.44\times10^7$ & $ 1.70\times10^{7}$& 0.102 & $960$& 21.15  & 0.981& 45.32& 20.23  \\
\hline  
\end{tabular}
\begin{tablenotes}
\item[a] A version of this halo simulated using the \fire-1 code was presented in \citet{Hopkins:2017}; uses a slightly different cosmology, box size, and starting redshift than the remainder of our simulations.
\item[b] A version of this halo simulated using the \fire-1 code was presented in \citet{Hopkins:2014} and \citet{Onorbe:2015}; uses a smaller box size than our other simulations.
\end{tablenotes}
 \end{threeparttable}
\end{table*}

Table~\ref{table:params} and Figure~\ref{fig:correlations} provide an overview of the galaxies in our simulation suite and their host dark matter halos. Table~\ref{table:params} includes information about the dark matter halos (columns 1 \& 2), the galaxies (columns 3-7), and the DMO versions of the halos (columns 8-10);  the entries in the table are ordered in terms of increasing $\mstar$ (column 3). Figure~\ref{fig:correlations} shows some basic properties of the galaxies in our suite: stellar half-mass radius $\rh$ (top), the ratio of dynamical mass to stellar mass within $\rh$ (middle, with dynamical mass being the sum of baryonic and dark matter mass), and one-dimensional stellar velocity dispersion $\sigma_{\star}$ (calculated as $\sigma_{\rm 3D,\star}/\sqrt{3}$ based on all of the stars within each galaxy; bottom) as a function of stellar mass. The galaxies from our suite are shown as cyan square symbols. For comparison, we also show data for low-mass galaxies in and around the Local Group from \citet{Kirby:2013,Kirby:2014} as gray circles (for satellites) and black circles (for non-satellites). 

The simulated galaxies from our suite agree well with observations for these basic properties of dwarfs $(\mstar - \rh - \sigma_{\star}-\mdyn)$. Although we do not focus on dynamics in this paper, we note in passing that the agreement in the bottom two panels indicates that rotational support in the stars must be minimal; otherwise, the measured dynamical mass would be significantly larger than that inferred from stellar kinematics. The stellar content of our simulated halos and its dependence on various properties of the halos are explored in detail in the following sections; our definition of the stellar mass associated with the central galaxy in each case is described in Appendix \ref{sec:appendixb}.

\subsection{Halo and Galaxy Assembly}

\label{subsec:assembly}

\begin{figure}
\includegraphics[width=\columnwidth]{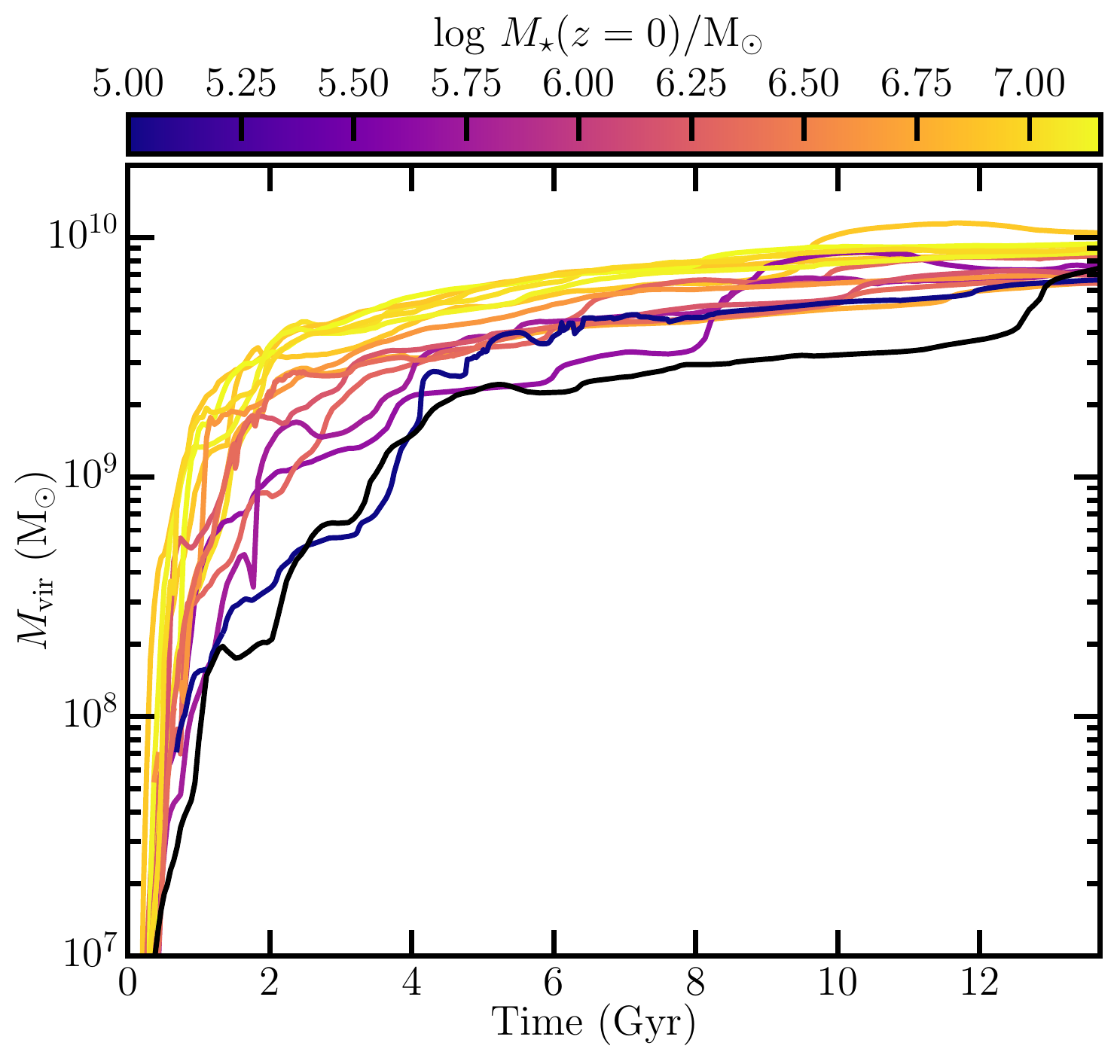}
    \vspace{-0.5cm}
  \caption{Dark matter halo mass assembly histories for our collection of halos. While all halos have $\mvir \approx 10^{10}\,\msun$ at $z=0$, their early evolution is varied, with scatter that exceeds 1 dex for $t \la 3\,{\rm Gyr}\;(z > 2)$. Each halo's line color indicates its stellar mass at $z=0$; halo m10a, which forms no stars, is plotted in black. There is an excellent correspondence between the virial mass at early times ($t \la 3$ Gyr or $z \ga 2$) and $\mstar(z=0)$.}
  \label{fig:mvir_vs_t}
\end{figure} 

\begin{figure}
\includegraphics[width=\columnwidth]{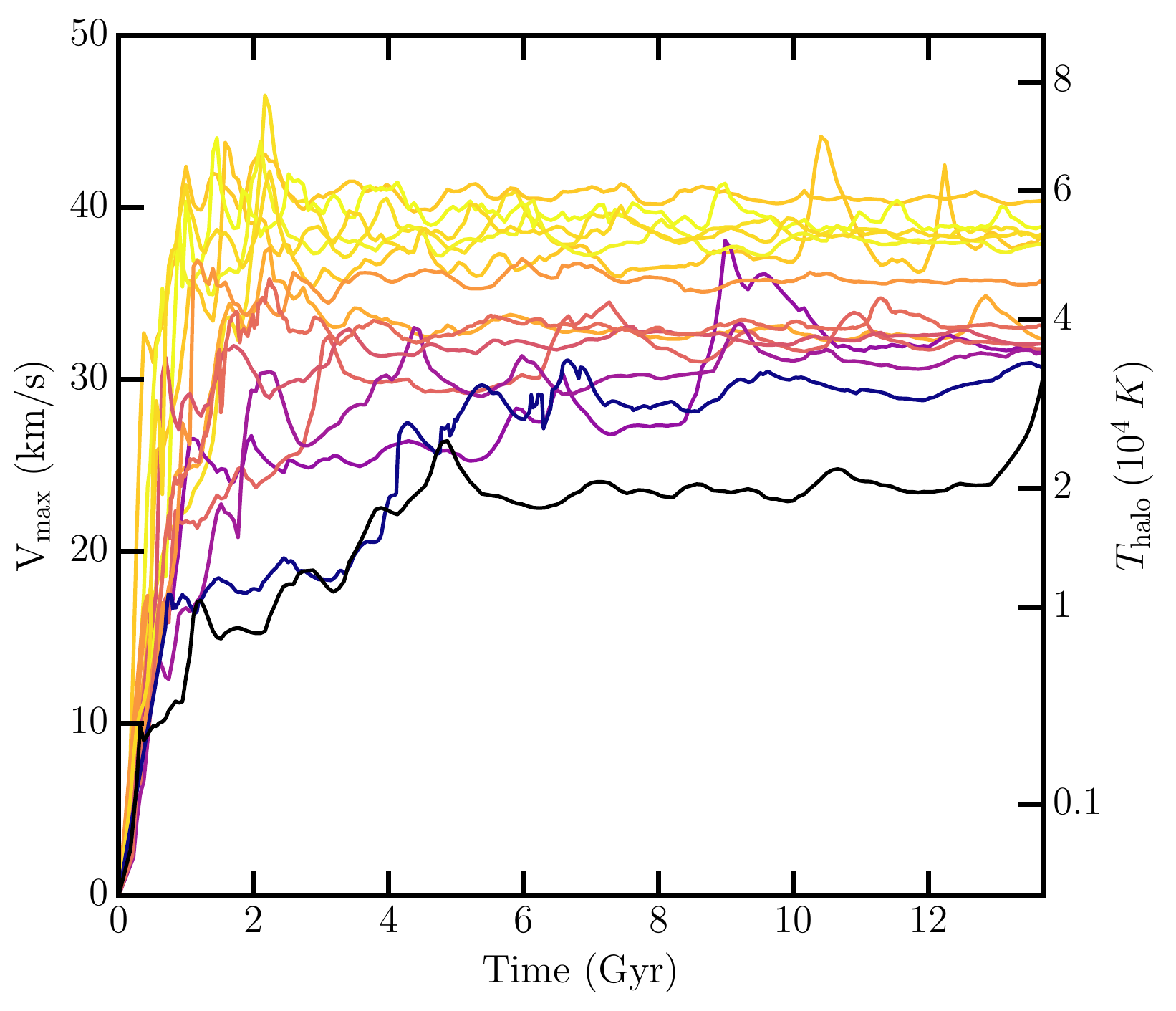}
    \vspace{-0.5cm}
  \caption{Maximum circular velocity $\vmax$ (and equivalent virial temperature $T_{\rm vir}$, on right axis) as a function of time along the main progenitor branch of each halo. As in Figure~\ref{fig:mvir_vs_t}, the line color indicates the stellar mass at $z=0$. The maximum circular velocity of each galaxy is typically set early ($t \sim 2$ Gyr or $z \sim 3$), and there is a strong correlation between $\mstar(z=0)$ and $\vmax$.}
  \label{fig:vmax_vs_t}
\end{figure}

Fig. \ref{fig:mvir_vs_t} shows the dark matter assembly histories for our halos; each line corresponds to one individual halo and is colored to reflect that galaxy's stellar mass at redshift zero. While, by design, all of the halos end up in a  narrow range around $\mvir(z=0)=10^{10}\,\msun$, there is significant spread in virial masses of their main progenitors at earlier times. From the coloring of the lines, and from Table~\ref{table:params}, it is clear that stellar mass at $z=0$ is strongly correlated with halo mass at early times ($t \approx 2-4$ Gyr, or $z \approx 3.1-1.7)$. This correlation persists, in slightly weakened form, to $z=0$. 

The evolution of $\vmax \equiv {\rm max}[{GM(<r)/r}]^{1/2}$ with time is shown in Figure~\ref{fig:vmax_vs_t}, again with colors indicating $\mstar(z=0)$. The correlation between $\vmax$ and $\mstar(z=0)$ is much stronger than that between $\mvir$ and $\mstar$ and is established early in the universe's history.
This is because $\vmax$ is a measure of the central gravitational potential and is set relatively early in a halo's growth history (as opposed to $\mvir$, which continues to grow even in the absence of physical accretion; \citealt{Diemer:2013,VandenBosch:2014}). Since our sample of halos spans a narrow range of $\mvir(z=0)$, higher $\vmax$ is indicative of higher concentration, which in turn points to earlier formation times. We see clear evidence of this correlation in Figure~\ref{fig:vmax_vs_t}, confirming the existence of a strong connection between a halo's central gravitational potential and its final stellar mass for our sample. 
\begin{figure}
\includegraphics[width=0.99\columnwidth]{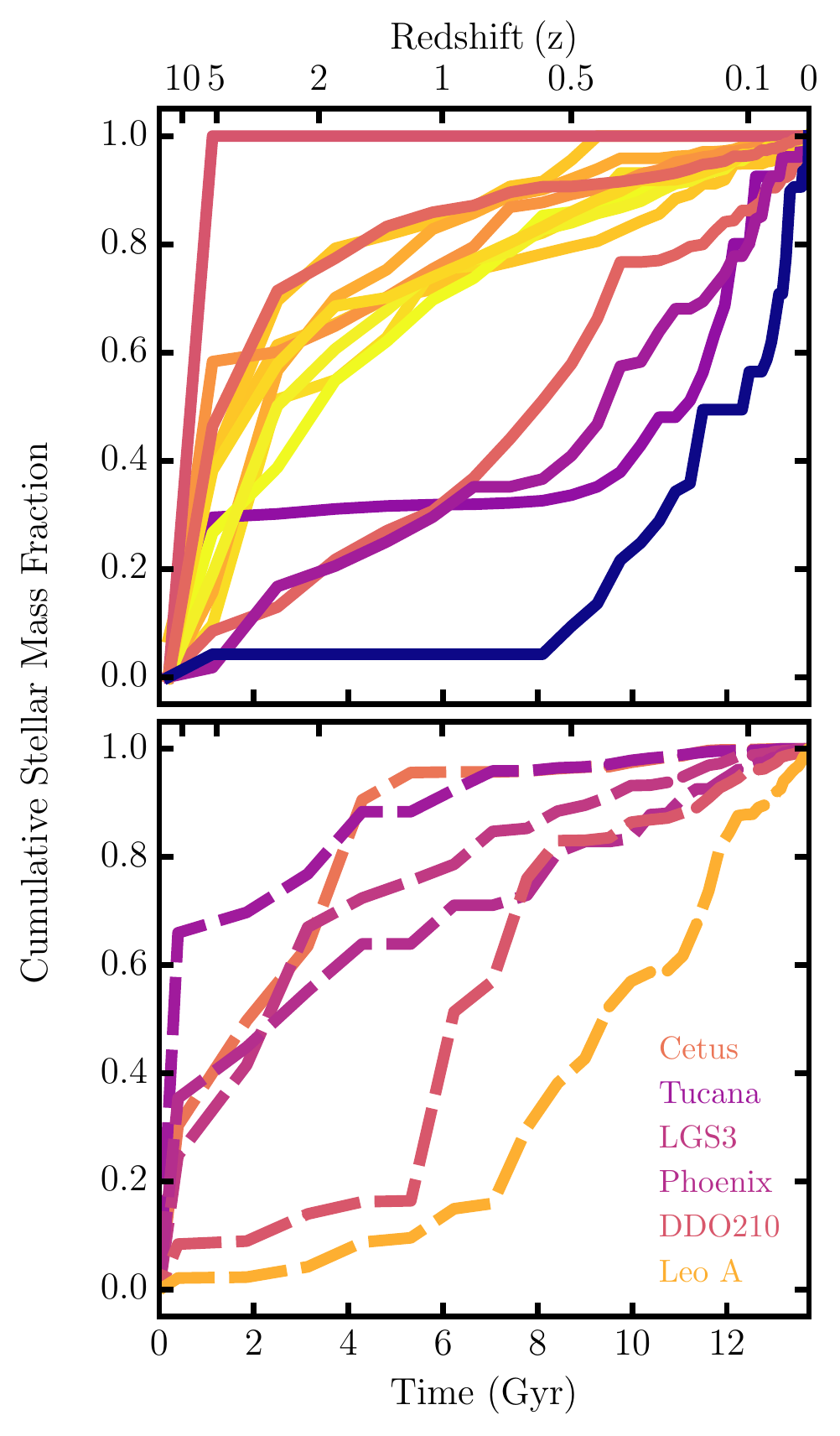}
  \caption{Star formation histories of simulated (top panel) and observed (bottom panel) dwarf galaxies. \textit{Top}: "archaeological" stellar mass assembly history for each galaxy, measured from the birth times of all of the stars in the galaxy at $z=0$ (mimicking SFHs derived from resolved star observations in the Local Group). \textit{Bottom}: SFHs based on resolved-star color-magnitude diagrams of observed Local Group field dwarfs with stellar masses similar to our simulated galaxies (from \citealt{Skillman:2014} and \citealt{Cole:2014}). The ordering in the legend follows the ordering of the lines at 5 Gyr and the color scale is identical to the simulated galaxies (i.e., using the same color scale as shown in Fig.~\ref{fig:mvir_vs_t}). Our simulated galaxies exhibit a variety of SFHs, similar to observations.}
  \label{fig:sfhs}
\end{figure} 
The equivalent scale for the virial temperature $T_{\rm vir}$ of each halo, where 
\begin{equation}
k\,T_{\rm vir} \equiv \frac{1}{2}\mu\,m_{\rm p}\,\vmax^2\,,
\end{equation}
is shown on the right-hand y-axis of Figure~\ref{fig:vmax_vs_t} (with $\mu=0.59$, appropriate for fully ionized gas with primordial composition; $m_{\rm p}$ is the proton mass). All of the halos in our suite have $3\times 10^{4} \leq T_{\rm vir} \leq 6\times 10^4\,{\rm K}$, with the exception of halo m10a (which has $T_{\rm vir} \approx 2\times 10^4\,{\rm K}$ until the very end of the simulation). Reionization heats the intergalactic medium to $T \approx 2\times 10^4\,{\rm K}$ \citep{Faucher-Giguere:2009,McQuinn:2016}, meaning that the gravitational potential of halo m10a is not sufficient to bind UV-heated gas in the post-reionization era. This halo also has significantly lower values of $\mvir$ and $\vmax$ than the rest of our sample until very recently, when it underwent a major (halo) merger. This unusual evolution of its gravitational potential with time explains why halo m10a does not form any stars, a point that is explored further in Sec.~\ref{subsec:reion}.

Figure~\ref{fig:sfhs} presents the star formation histories (SFHs) of our galaxies (top panel), along with a comparison to measured SFHs of Local Group galaxies in the same mass range based on resolved color-magnitude diagram (CMD) analyses\footnote{\!Some of the data from \citet{Skillman:2014} were first analyzed in \citet{Cole:2007}, \citet{Hidalgo:2009, Hidalgo:2011}, and \citet{Monelli:2010a, Monelli:2010b}.} in \citet{Skillman:2014} and \citet{Cole:2014}. While it is common in simulation-based studies to consider the main branch SFH (i.e., the star formation rate in the main progenitor of the halo at each time), observational SFHs from CMD studies are inherently "archaeological": all of the stars present at $z=0$ are used to calculate when a given fraction of the present-day stars were formed, irrespective of the distribution of stars over all progenitors at a given time. We therefore compute archaeological SFHs for our simulated galaxies as well, and in both cases, plot the fraction of stellar mass at $z=0$ formed by a given cosmic time (or redshift). Encouragingly, our simulations exhibit a similar diversity of SFHs as is observed.

While the majority of the galaxies in our sample form over 50\% of their stars at early times (by $t\approx 4$ Gyr or $z \approx 1.7$), there are also galaxies that form stars at a nearly constant rate (averaged over $\sim 500$ Myr time-scales) or that have dominant star formation at late times. Two galaxies even "self-quench" (i.e., they stop forming stars owing to internal feedback processes), one at $z \sim 5$ and one at $z \sim 0.5$. We return to the question of self-quenching in Section~\ref{subsec:reion}.

\begin{figure*}
\includegraphics[width=1.0\textwidth]{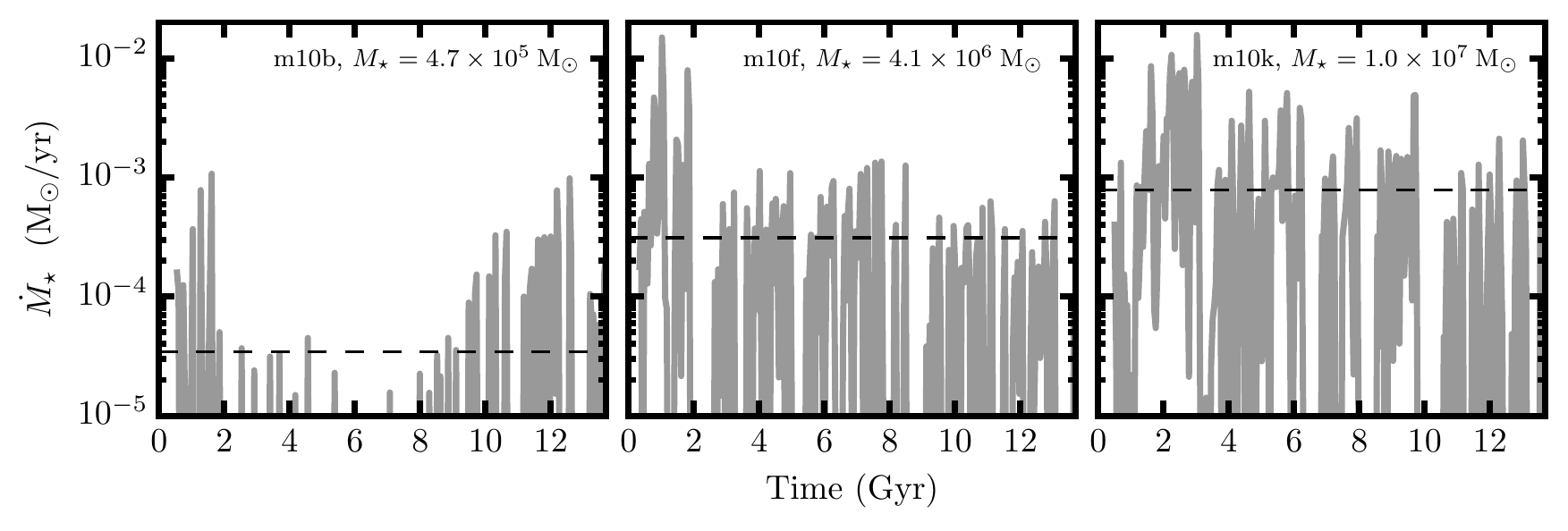}
  \caption{Star formation rate, averaged over 50 Myr time intervals, as a function of time for a low-mass galaxy (left; $M_{\star}=4.7\times10^5$ $M_\odot$), an intermediate-mass galaxy (center; $M_{\star}=4.1\times10^6$) and a high-mass galaxy (halo m10k, $M_{\star}=1.0\times10^7$ $M_\odot$) from our simulated sample. Dashed horizontal lines show the average star formation rate for each galaxy over the age of the Universe. Galaxies with higher stellar mass at $z=0$ have higher star formation rates, which in turn drive larger gravitational potential fluctuations. Star formation in all of the galaxies is bursty, with significant variations around the mean.}
  \label{fig:sfr_vs_t}
\end{figure*}
\begin{figure*}
\includegraphics[width=1.0\textwidth]{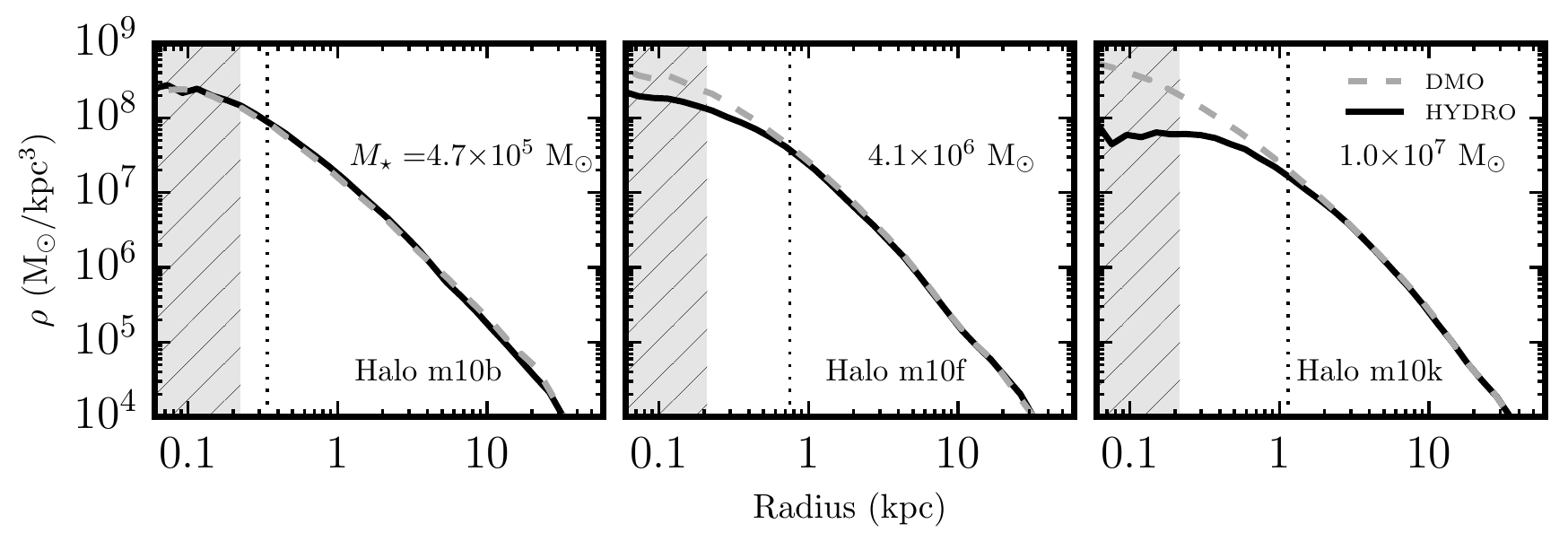}
  \caption{Density profiles for the three halos plotted in Figure~\ref{fig:sfr_vs_t}. Dotted vertical lines mark the galaxy half-mass radius in each case. The gray hatched region shows where numerical relaxation may affect the density profiles according to the Power et al. criterion. The dashed gray (solid black) line corresponds to the density profile for the DMO (hydrodynamical) run for each halo. The amount of central density reduction and size of any core produced is proportional to the stellar mass of the galaxy.}
  \label{fig:density_trippro}
\end{figure*}
Figure~\ref{fig:sfr_vs_t} shows the archaeologically-determined star formation rate as a function of time for the three halos shown in Figure~\ref{fig:density_trippro}, highlighting the degree of burstiness in each case (the mean star formation rate is shown as a dashed horizontal line in each panel). The star formation rates are averaged over 50 Myr periods, which is much finer resolution than can be obtained from observations for most ages. The star formation histories are clearly bursty on 50 Myr timescales, with fluctuations that can exceed a factor of 100 in adjacent bins (see also \citealt{Stinson:2007,Ricotti:2008,Shen:2014,Dominguez:2015,Boylan-Kolchin:2015,Sparre:2015}). The star formation rate correlates with stellar mass throughout the simulation -- i.e., galaxies with higher $z=0$ stellar masses have higher star formation rates at essentially every epoch -- implying that galaxies with larger stellar masses at $z=0$ have likely experienced larger feedback-driven outflows (see also Section~\ref{subsec:fb}).

\subsection{Central Densities} 
\label{subsec:central_density}
One of the most pressing questions in galaxy formation (and dark matter physics) is how the centers of dwarf dark matter halos are affected by galaxy formation. While it had long been assumed that the high dynamical mass-to-light ratios measured for Local Group dwarfs pointed to a relative unimportance of baryons for shaping the dark matter structure of galaxies, recent numerical and analytical work has established that baryons may indeed play a crucial role in setting the structure of dark matter halos even in faint ($\mstar \sim 10^7\,\msun$) galaxies that are dominated by dark matter at their centers \citep{Zolotov:2012,Pontzen:2012,DiCintio:2014,Onorbe:2015,Chan:2015,Read:2016a,Tollet:2016}. With our sample, we can explore the connections among star formation, halo assembly, and dark matter structural changes for halos with $\mvir(z=0) \approx 10^{10}\,\msun$.

In Figure~\ref{fig:density_trippro}, we show the density profiles for the DMO (dashed gray curves) and hydrodynamical (solid black curves) versions of three halos. The stellar content of the halos increases left to right, and there is a clear trend of greater central density reduction between DMO and hydrodynamical simulations as the stellar mass at $z=0$ increases. Our galaxies with $\mstar < 2 \times 10^{6}\,\msun$ do not have any appreciable reduction in central density. $\mstar=2 \times 10^{6}\,\msun$ (or $\mstar/\mvir = 2 \times 10^{-4}$) appears to be a critical stellar mass at this halo mass: galaxies with higher stellar mass can affect the density distribution of their host halos, while galaxies with lower stellar mass cannot. 

\begin{figure}
\includegraphics[width=\columnwidth]{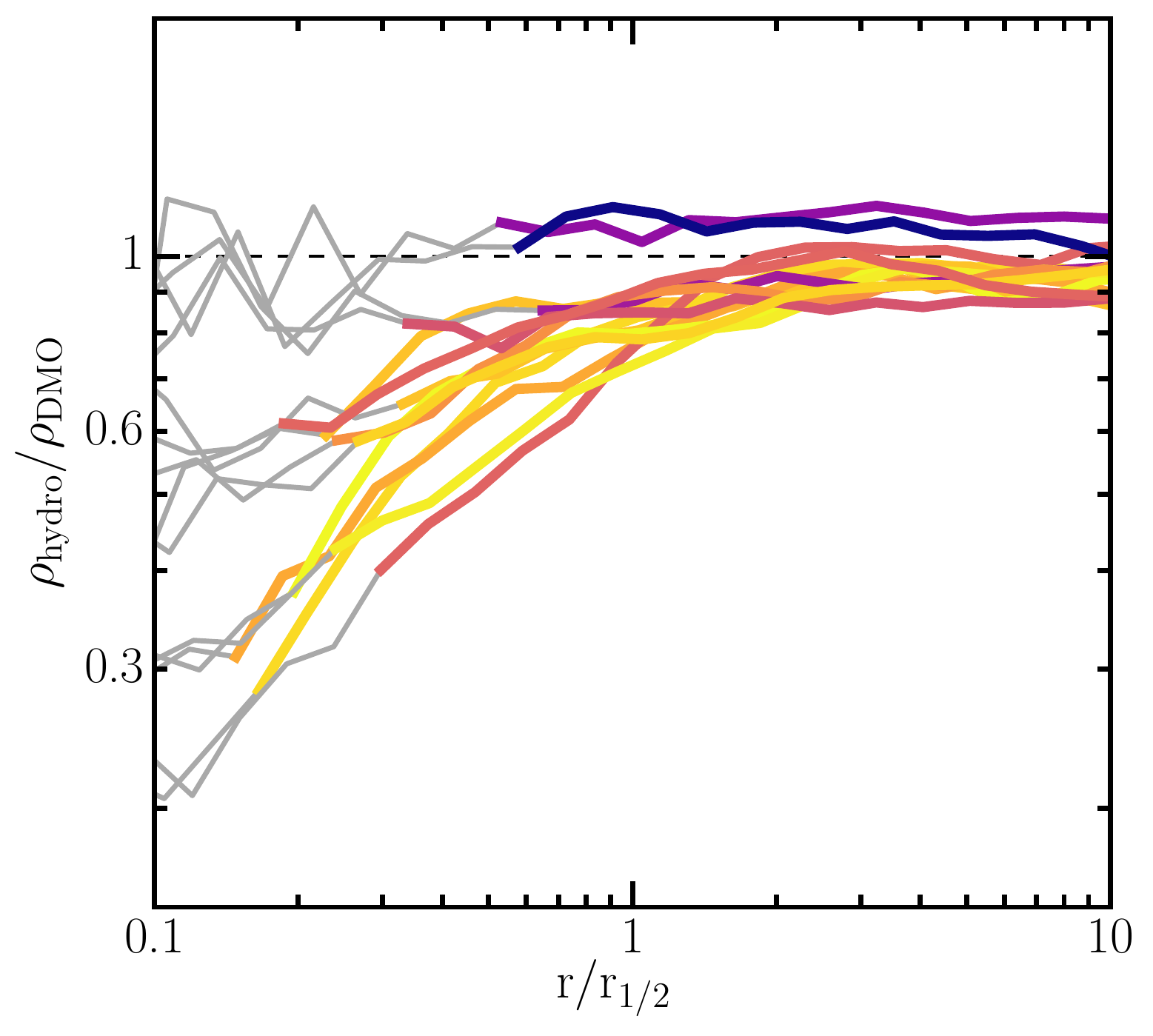}
\caption{The effects of star formation and feedback on central dark matter density. Each curve shows the ratio of a halo's density profile obtained in the hydrodynamic simulation to its DMO counterpart as a function of radius (scaled by the galaxy's half-mass radius). As in previous plots, color corresponds to $\mstar(z=0)$, and lines are plotted in gray for radii smaller than the Power radius. At large distances from the halo center, all density profiles have amplitudes that are within 10\% of their DMO counterparts. Within the half-mass radius, however, the central density can be reduced in the hydrodynamic runs, with the amount of reduction strongly correlated to $\mstar(z=0)$: the lowest-mass galaxy is virtually unchanged from the DMO run, while the highest-mass galaxies have large reductions in central density.}
\label{fig:density_ratiopro}
\end{figure}

To better understand the modification of the central dark matter structure in our simulated sample, Fig.~\ref{fig:density_ratiopro} shows the ratio of each galaxy's density profile in the hydrodynamical run to $\rho(r)$ obtained from its DMO version. The horizontal axis is scaled by the galaxy half-mass radius, $r_{1/2}$. The density profile ratios are colored by $\mstar(z=0)$, identically to previous figures; below the Power (\citeyear{Power:2003}) radius, the line coloring is changed to gray. A number of interesting trends appear in the Figure. On large scales ($r \gg r_{1/2}$), the amplitude of the $\rho_{\rm hydro}(r)$ is very similar to $\rho_{\rm DMO}(r)$, indicating that baryonic physics has minimal effects there. On small scales ($r \la r_{1/2}$), however, the density profiles in many runs are systematically lower in the hydrodynamical simulations relative to the DMO simulations, pointing to the efficacy of stellar feedback at modifying the central gravitational potential even in dwarf galaxy halos. It is also interesting to note that the size of this effect depends systematically on stellar mass, echoing the results shown in Figure~\ref{fig:density_trippro}. The galaxies with the lowest $\mstar$ (darkest curves) show the least central density reduction -- including no reduction at all for 2 of the systems -- while the highest $\mstar$ galaxies show the largest central density reduction. Furthermore, $r_{1/2}$ is an excellent indicator of the radial scale at which any density modification occurs. Our simulations therefore predict that the density profiles of low-mass dwarf galaxies in \lcdm\ should be virtually unmodified (relative to DMO predictions) on scales larger than $r_{1/2}$. The clear trend seen in central density reduction with respect to stellar mass formed (at fixed mass resolution), coupled with our extensive tests of numerical convergence (Appendix~\ref{sec:appendixa} and \hopkins), strongly point to a physical, not numerical, origin.

\begin{figure}
\includegraphics[width=\columnwidth]{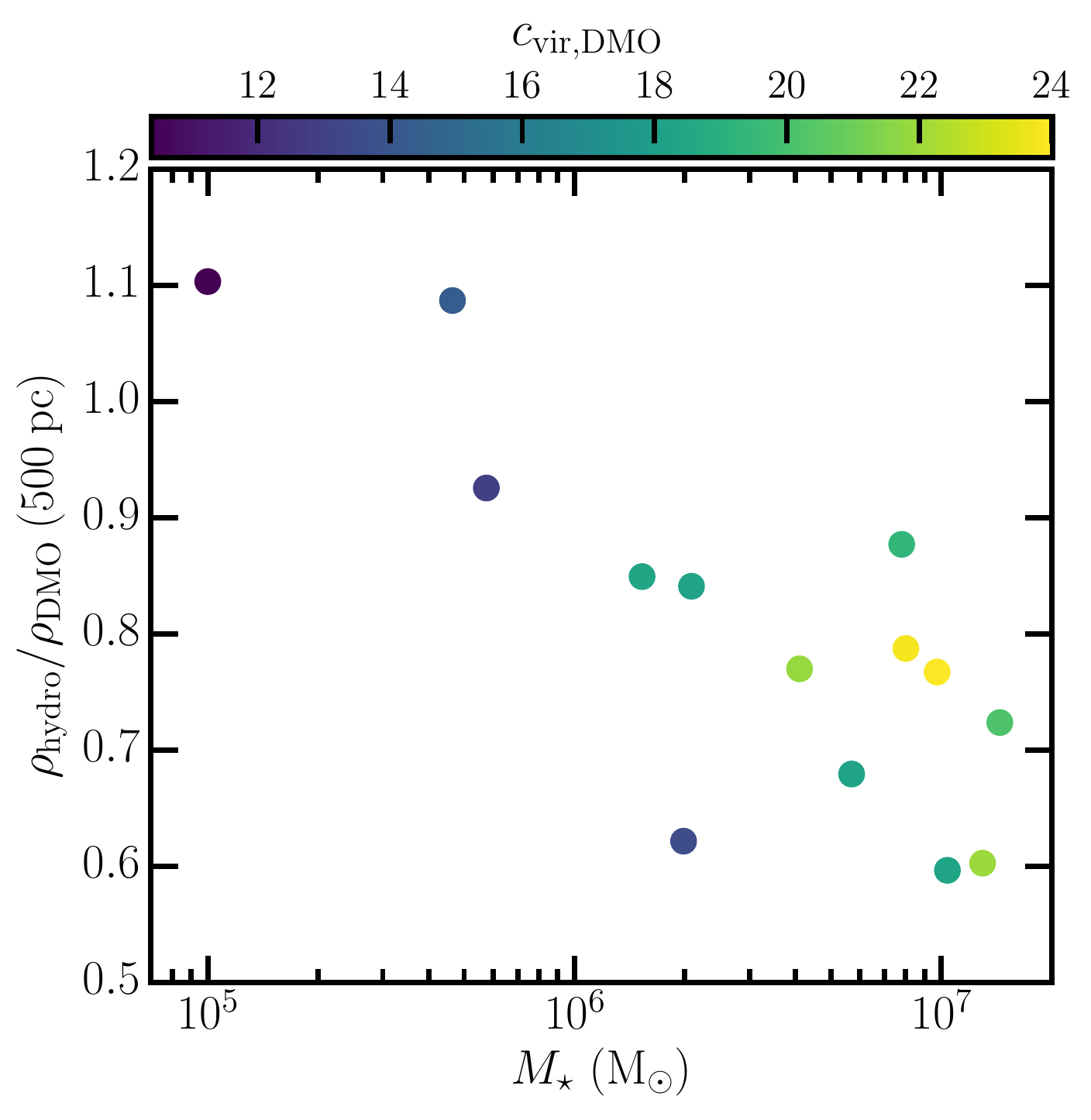}
    \vspace{-0.5cm}
  \caption{The correlation between $M_{\star}$ and the central density reduction in the hydrodynamic runs relative to the DMO runs at 500 pc. There is little to no reduction in central density below $10^{6}\,\msun$, while more massive systems see significant reduction. This figure offers a complementary view of Figure~\ref{fig:density_ratiopro}, in which density reduction is shown as a function of radius scaled by $\rh$: it shows the density reduction at a fixed \textit{physical} radius. The coloring of the points indicates the concentration of each galaxy's halo in the DMO run.}
  \label{fig:density_mstar}
\end{figure}
As an alternate way of looking at the central density reduction as a function of stellar mass, Figure~\ref{fig:density_mstar} shows the ratio of density in the hydrodynamical run to the DMO run for each halo at a fixed physical radius of 500 pc (as opposed to Figure~\ref{fig:density_ratiopro}, which shows central density reduction as a function of $r/\rh$). The density reduction at a fixed physical radius also shows a clear correlation with $\mstar$. The coloring of the points in Figure~\ref{fig:density_mstar} indicates the concentration parameter of each halo measured in the DMO run. Even with our relatively large suite of galaxies at fixed halo mass, it is difficult to discern if there is a trend in density reduction with halo concentration at fixed $\mstar$.

\begin{figure*}
\includegraphics[width=1.0\textwidth]{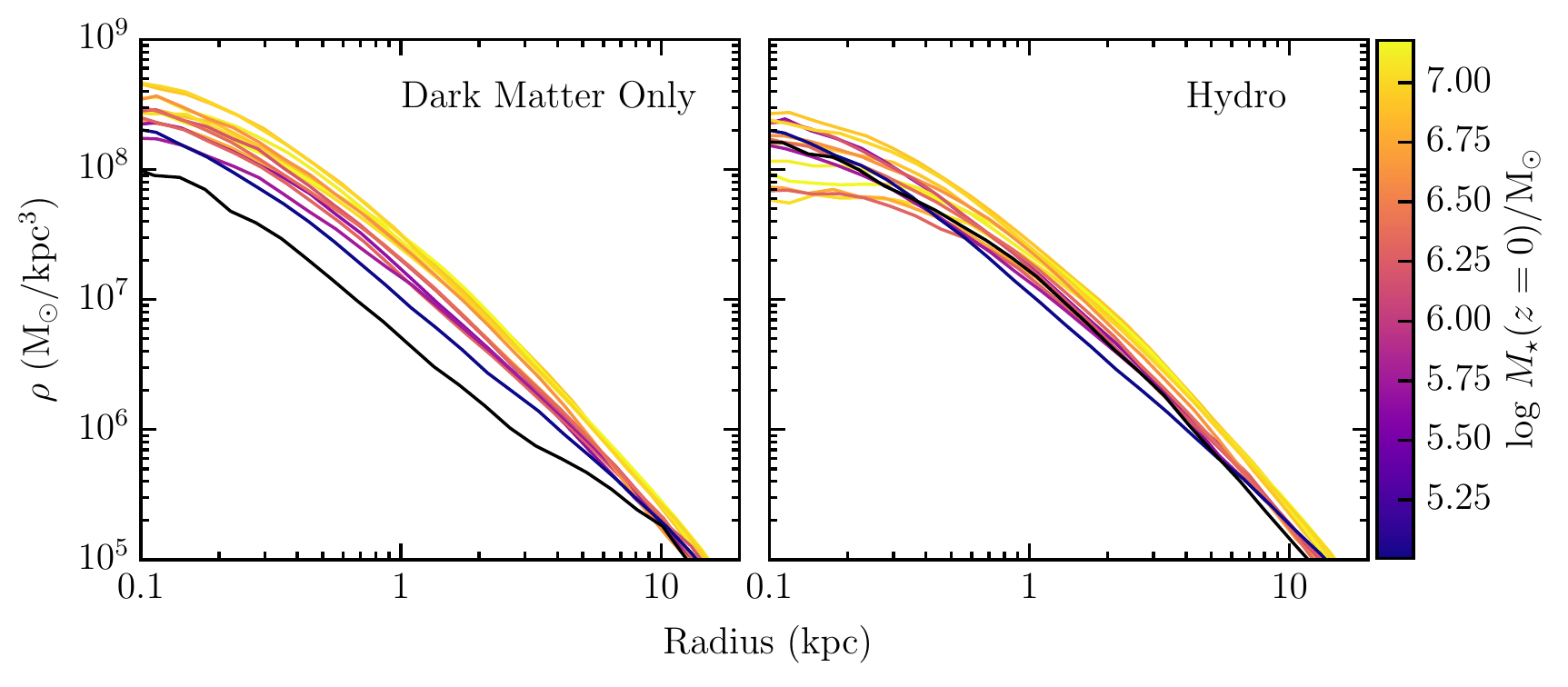}
  \caption{\textit{Left}: density profiles from DMO runs. \textit{Right}: density profiles from full galaxy formation physics runs. In both cases, the line coloring indicates the stellar masses of the galaxies in the hydrodynamical runs. There is a strong correlation between central density in the DMO run and the stellar mass formed in the hydrodynamical version for each halo. DMO density may therefore explain the scatter observed in stellar mass within the narrow range of $z=0$ halo masses simulated here.}
  \label{fig:density_dblpro}
\end{figure*}

Figures~\ref{fig:density_trippro}--\ref{fig:density_mstar} demonstrate that more massive (dwarf) galaxies have greater reduction in the central densities of their host halos and that the radial scale of this central density reduction is set by the size of the galaxy. In Figure~\ref{fig:density_dblpro}, we show that density amplitudes \textit{in the dark-matter-only simulations} are excellent predictors of stellar mass. The right panel of the figure shows density profiles in the hydrodynamical run, with line color again mapped to stellar mass at $z=0$. At large scales ($r \gg 1\:\mathrm{kpc}$), the densest halos are also the ones that form the most stars. In the centers of these halos, however, the central density reduction wipes out any trace of this correlation. The left panel of the Figure shows the density profiles in the DMO runs, with colors indicating the stellar mass in the hydro version of each run at $z=0$. Amazingly, the correlation between density and stellar mass exists at essentially all radii in the DMO run: the stellar mass of a halo at fixed $\mvir=10^{10}\,\msun$ can be predicted directly from the central density (or $\vmax$, or formation time) of that halo in a DMO simulation. This intriguing result reinforces trends identified in Section~\ref{subsec:assembly}.

\begin{figure}
\includegraphics[width=\columnwidth]{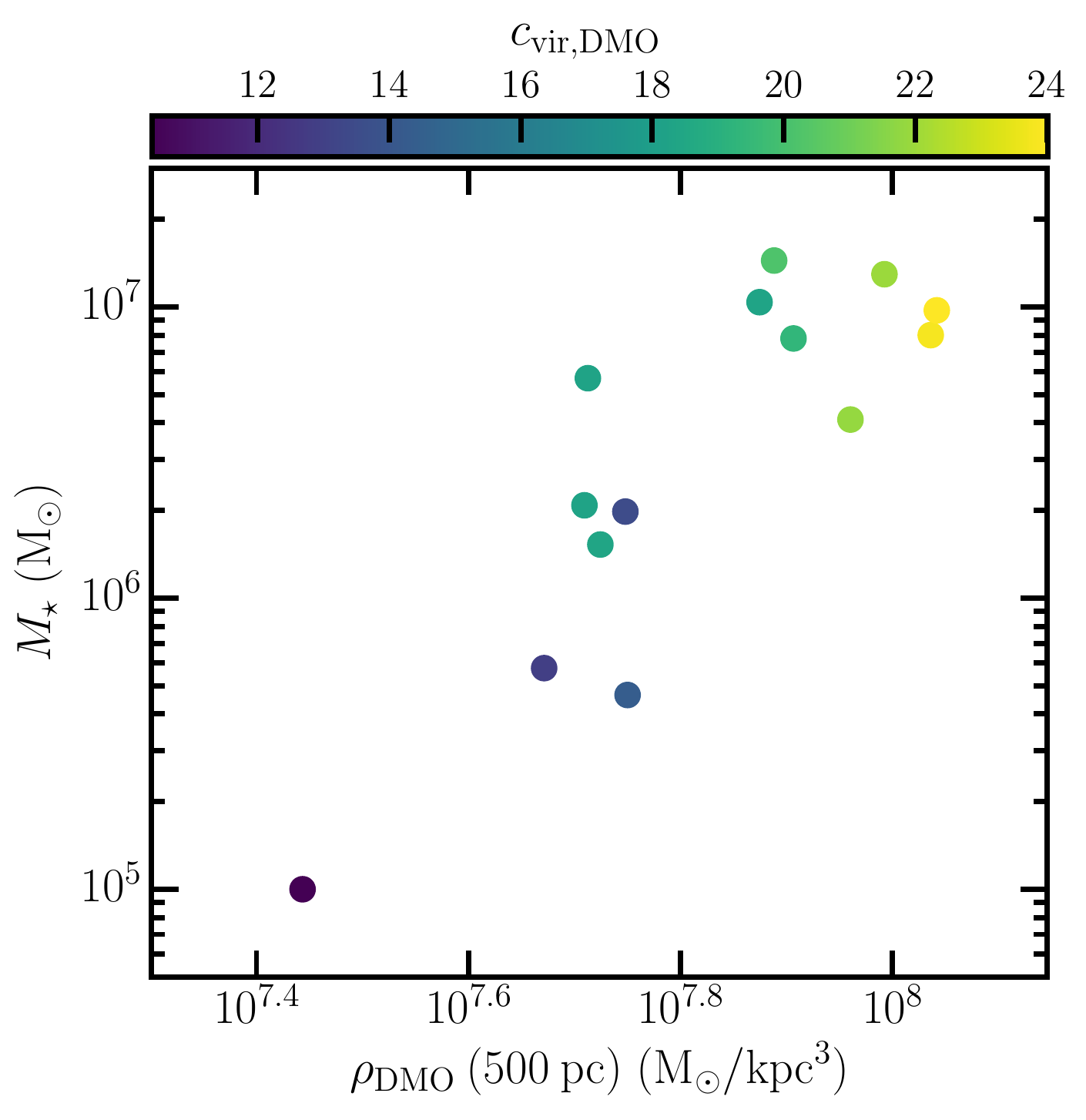}
    \vspace{-0.5cm}
  \caption{Relationship between stellar mass and central dark matter density in the DMO runs. There is a clear correlation between the stellar mass formed and the amplitude of the density profile at 500 pc in the DMO version of each simulation; this figure is equivalent to taking a slice through the left panel of Figure~\ref{fig:density_dblpro} at 500 pc. The coloring of each point shows the measured concentration in the DMO run.
}
  \label{fig:rho_dmo_vs_mstar}
\end{figure}

This correlation is explored further in Figure~\ref{fig:rho_dmo_vs_mstar}, which plots the stellar mass of each galaxy as a function of the amplitude of the DMO density profile at 500 pc. The connection between the two is apparent and points to halo density as a "second parameter" in abundance matching that determines the scatter in $\mstar$ at fixed $\mvir$. A more detailed exploration of the connection between halo density, halo mass, and stellar mass across a wider range of simulated halo masses is clearly warranted and will be presented in a future \fire-2 paper. For now, we note that the scatter obtained in our simulations is consistent with $\pm 0.5\,{\rm dex}$ or so, larger than is found for more massive systems (e.g., \citealt{Behroozi:2013b}) but not consistent with completely stochastic galaxy formation at these masses (see, e.g., \citealt{Garrison-Kimmel:2016}).

\section{Discussion} 
\label{sec:discussion}
\subsection{Effects of reionization}
\label{subsec:reion}
\begin{figure*}
\includegraphics[width=1.0\textwidth]{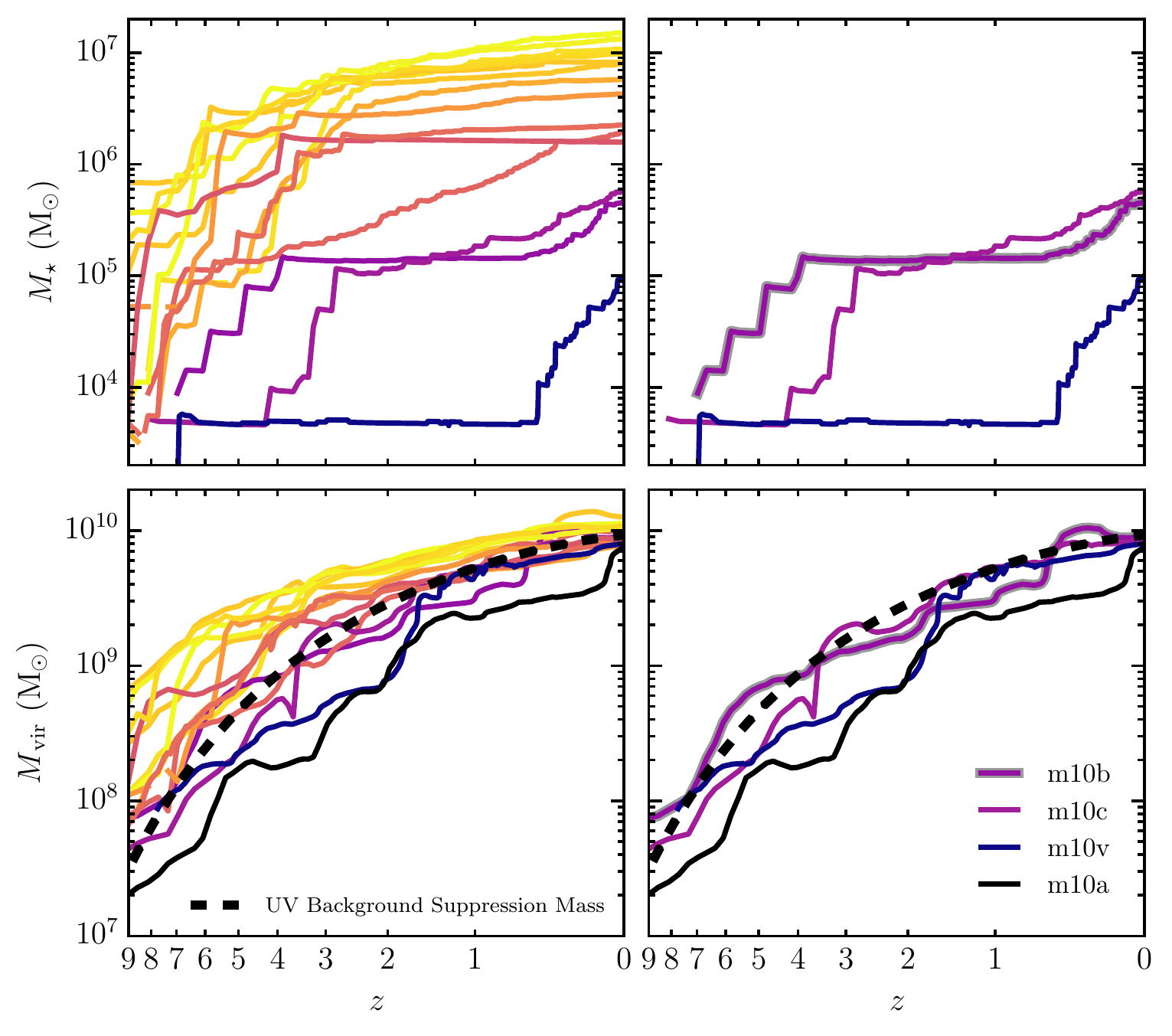}
    \vspace{-0.5cm}
  \caption{\textit{Left}: stellar (upper panels) and dark matter (lower panels) mass growth along the main progenitor branch of the halos in our suite plotted as a function of redshift. The black dashed line is the characteristic mass ($M_{\mathrm{char}}$) at which halos have lost half of their baryons owing to the UV background (from \citealt{Okamoto:2008}). \textit{Right}: same as left panels, but focusing on the systems with the lowest stellar masses. Halo m10a (black line), which forms no stars, falls below $M_{\mathrm{char}}$ at all redshifts, while the three other halos that have the lowest values of $\mstar(z=0)$ fall below $M_{\mathrm{char}}$ for extended periods. These results highlight the importance of reionization feedback in setting the stellar content of these systems.}
  \label{fig:mvir_vs_z}
\end{figure*}

The results of the previous section demonstrate that there are tight connections between the dark matter structure of $\mvir(z=0)\approx 10^{10}\,\msun$ halos in DMO simulations, halo assembly, galaxy stellar mass, and the central density structure of galaxies. Earlier-forming, more concentrated, higher $\vmax$ halos form more stars than do later-forming, less concentrated, lower $\vmax$ halos. In fact, the lowest $\vmax$ and lowest concentration halo in our sample, m10a, does not form \textit{any} stars, despite having a $z=0$ virial mass that is only 5-15\% lower than halos that form up to $6\times 10^{6}\,\msun$ of stars. This seemingly puzzling behavior is linked to its late-time assembly, as seen in Figure~\ref{fig:vmax_vs_t}, and the halo mass-dependent effects of cosmic reionization. Dwarfs with reionization-era virial temperatures below $\sim10^4$ K (such as m10a) are particularly susceptible to the effects of reionization-induced feedback.

The relationship between stellar mass growth and halo growth for our suite is explored in more detail in Figure~\ref{fig:mvir_vs_z}. The left panels show all of the halos, while the right panels focus on those with the lowest stellar masses at $z=0$ (halos m10a, m10b, m10c, and m10v). The lower panels show the dark matter mass assembly history, while the upper panels show the growth of $\mstar(<\rvir)$ as a function of redshift. Halo m10a, plotted in black, is noticeably lower in both virial mass and $\vmax$ (see fig. \ref{fig:vmax_vs_t}) compared to the other halos for the first 8 Gyr of cosmic time. The effects of a late major merger are also visible in these plots, as the virial mass jumps just before $z=0$. Since $\vmax$ is more related to the central mass distribution than to mass at large radii (which is probed by $\mvir$), there is a delay of a crossing time ($\approx1$ Gyr) before $\vmax$ is affected by the merger. Even though we expect the vast majority of halos at $\mhalo \approx 10^{10}\,\msun$ to host galaxies \citep{Sawala:2013,Benitez-Llambay:2016}, the diverse assembly histories of such halos mean that some can be completely dark at $z=0$. However, most \fire\ halos with masses that are a factor of $\sim 5$ lower at $z=0$ \textit{do} form stars \citep{Wheeler:2015a}; halo m10a is the only target halo studied at high resolution that has failed to form stars within the main \fire\ sample.

To see the expected sensitivity of halo m10a to reionization, we plot the
characteristic reionization suppression mass $M_{\rm char}(z)$ from
\citet{Okamoto:2008} as a black dashed line in Figure~\ref{fig:mvir_vs_z}. While
this suppression mass is usually defined as the mass at which a typical halo has
a baryon fraction that is suppressed by 50\% relative to $f_{\rm b}$ owing
solely to UV background feedback
(\citealt{Gnedin:2000,Dijkstra:2004,Hoeft:2006,Okamoto:2008}; see also
\citealt{FaucherGiguere:2011}), which is somewhat arbitrary, the suppression is
relatively abrupt in mass: the mass corresponding to 30\% or 70\% suppression is
close to $M_{\rm char}$ (see also \citealt{Noh:2014}), which makes the threshold
more physically meaningful.  Halo m10a lies below $M_{\rm char}$ at all
redshifts, indicating it never was able to accrete and retain enough baryons to
accumulate the cold gas necessary for star formation. This halo has the lowest
baryon fraction (5\% of the cosmic baryon fraction) at $z=0$
(Table~\ref{table:params}). The lower right panel of Figure~\ref{fig:mvir_vs_z}
also shows that three other halos have masses falling below $M_{\rm char}(z)$
for significant periods of time: halo m10v does so from $z\approx 6.5-1$, halo
m10c for $z \ga 3.5$, and halo m10b for $4 \ga z \ga 0.7$. While none of these
three systems are quenched at $z=0$, all show signs of reionization suppression
in their SFHs (Fig.~\ref{fig:sfhs}).

\begin{figure*}
\includegraphics[width=0.49\textwidth]{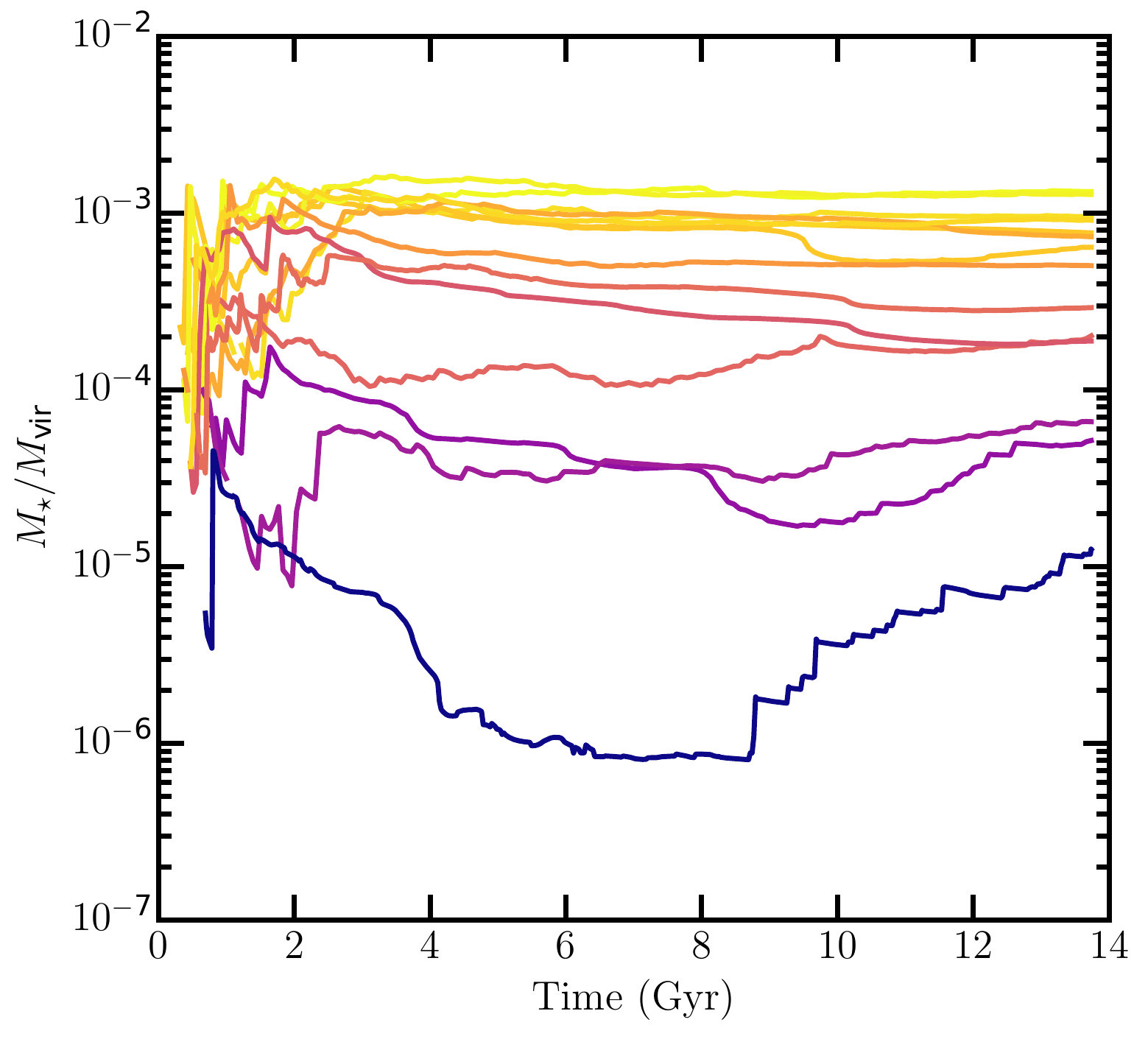}
\includegraphics[width=0.49\textwidth]{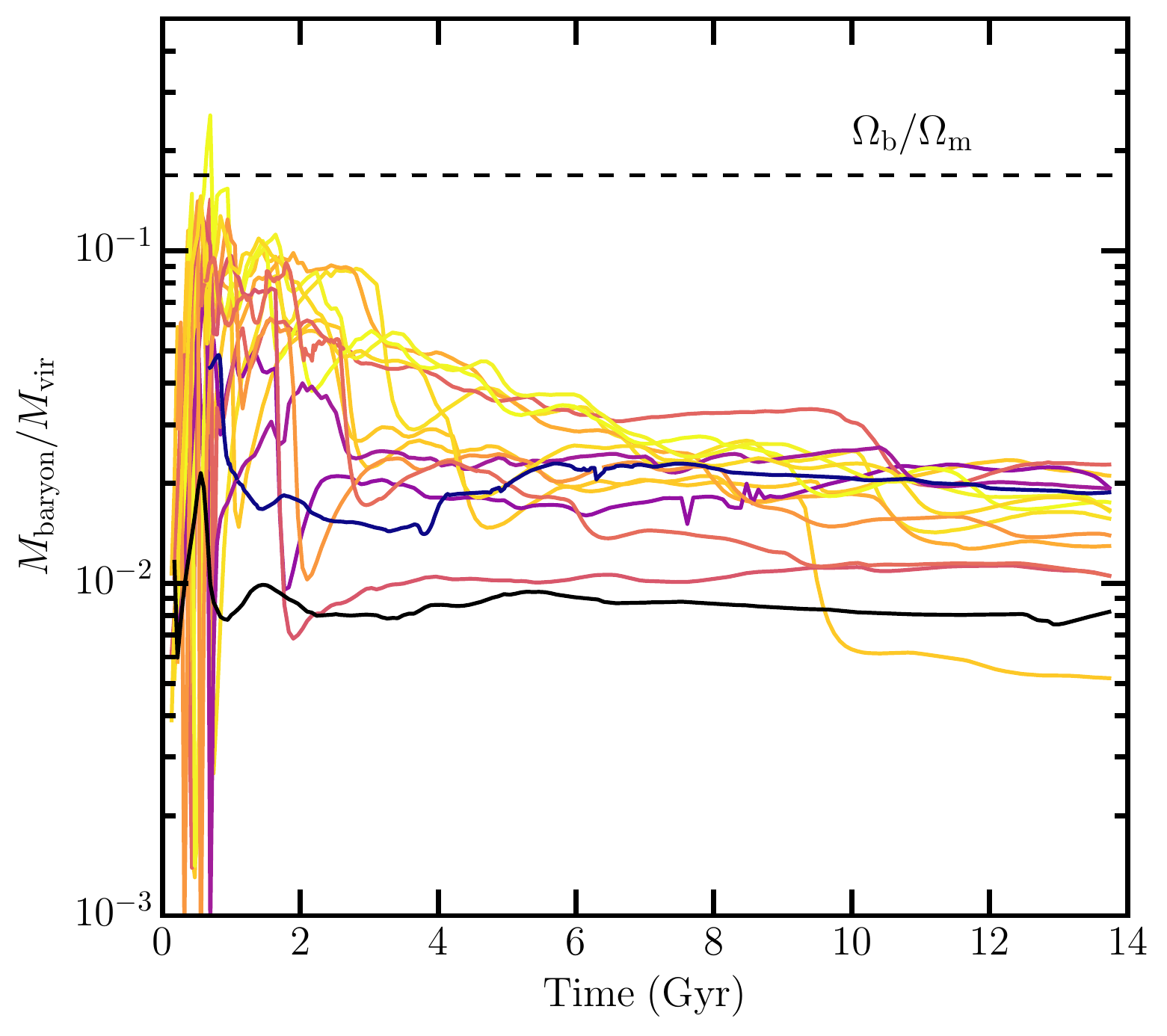}
  \caption{\textit{Left}: The ratio of stellar mass (in the central galaxy) to virial mass for each halo as a function of time. \textit{Right}: The baryon fraction within $\rvir$ for each of the 15 halos measured throughout time. The dashed black horizontal line represents the cosmic baryon fraction. Galaxies with higher $\mstar(z=0)$ exhibit a steady decrease of baryons, while lower $\mstar$ galaxies lose their baryons earlier (likely a result of their shallower gravitational potentials / lower values of $\vmax$).}
  \label{fig:baryon_frac}
\end{figure*} 

Halo m10v, which was below the suppression threshold for most of the first half of cosmic history, exhibits a complete lack of star formation (after a small initial burst) for the corresponding period in Figure~\ref{fig:sfhs}. Only after a late-time merger brings its mass above the suppression threshold does it begin to form stars in a more sustained manner. Halo m10b sees its star formation suppressed substantially until its virial mass exceeds $M_{\rm char}$. The corresponding SFH shows an early burst with a long pause at intermediate ages, which may be a signature of feedback from cosmic reionization \citep{Ricotti:2009,Weisz:2014b, Benitez-Llambay:2015}. This behavior is consistent with the observed SFH of Leo T \citep{Clementini:2012, Weisz:2014a}, perhaps indicating that Leo T underwent a major halo merger at $z \sim 2$ (see \citealt{Ricotti:2009} for a somewhat different scenario along these same lines). Halo m10c is able to form stars for much of its early evolution; it is suppressed after falling below $M_{\rm char}(z)$ and exhausting its cold gas supply, only to re-emerge as a star-former once it gains sufficient mass. 

Two galaxies, m10d and m10i, show the interesting behavior of self-quenching at $z=0$, even though their host halos are well above the reionization suppression threshold. As can be seen from Table~\ref{table:params}, these halos have no cold gas at $z=0$, indicating this self-quenching is likely to be long-lived. However, strong blowouts of gas caused by a large number of SNe going off concurrently are immediate precursors to both of these self-quenching events. The timing of these SNe blasts is stochastic, and resimulations of the same dwarfs do not always produce identical behavior. From the small fraction of resimulations that did show this effect, it was not entirely clear how statistically robust it is to expect self quenching at this mass scale. At slightly higher masses, quenched field galaxies are either extremely rare or exceedingly hard to detect at cosmological distances \citep{Geha:2012}. While the self-quenched galaxies in our sample raise the intriguing possibility of a larger population of self-quenched halos at low stellar masses (see also \citealt{Wetzel:2015,Weisz:2015,Fillingham:2015,Wheeler:2015a}), a more dedicated look at this effect is necessary before it is possible to make specific predictions.

A handful of quenched dwarf spheroidals with $\mstar \sim 10^6-10^7\,\msun$ are known to exist in the "field", including Cetus and Tucana in the Local Group and KKR 25 \citep{Karachentsev:2001} and KKs 3 \citep{Karachenstev:2015} at somewhat greater distance. While the lack of star formation in Cetus and Tucana is often interpreted as evidence that they were once within the virial radius of the Milky Way or M31 (e.g., \citealt{Teyssier:2012}), our work raises the possibility that they were not quenched through interaction with a more massive halo (see \citealt{Benitez-Llambay:2013} for an alternative explanation of quenching without requiring interaction with a Local Group giant). Given the sensitivity of halos at the $\mhalo \sim 10^{10}\,\msun$ mass scale to reionization, it will be important to explore more broadly the effects of different UV background models (e.g., \citealt{Onorbe:2016}) on the SFHs of simulated $\mstar \sim 10^6\,\msun$ galaxies (Elbert et al., in preparation). This exploration will be relevant for understanding how the timing of reionization impacts our results, as the UV background model adopted here has a reionization redshift ($z\approx 10$) that is slightly higher than what is derived in the latest Planck results ($z\approx 9$; \citealt{Planck:2016b}).

\subsection{Baryon fractions and halo masses}
\label{subsec:fb}
The time evolution of $\mstar/\mvir$ (left) and $M_{\rm baryon}/\mvir$ (right), calculated for the main progenitor at $z>0$, are shown in Figure~\ref{fig:baryon_frac}. The halos that form $\ga 5\times10^6\,\msun$ of stars have stellar-to-virial mass ratios that remain nearly constant or decline slowly through time, indicating that their star formation rates (plus contributions from mergers) closely match the halo growth rates. Their total baryon fractions typically show secular declines with time, however, pointing to a slow loss of baryons. Lower $\mstar$ systems show larger variations in $\mstar/\mvir$ but nearly constant values of $M_{\rm baryon}/\mvir$, meaning they are \textit{not} losing baryons after an initial period of rapid baryon loss. This difference likely is caused by the higher star formation rates in the higher $\mstar$ systems, which leads to somewhat stronger outflows. At the halo masses we are considering here, with equivalent virial temperatures of $\sim 4\times 10^4\,{\rm K}$, slight changes in heating rates can be the difference between baryons evaporating from the halo and baryons remaining in a tenuous, diffuse phase at large distances. These effects will be explored in more detail in Fitts et al. (in preparation).

A particularly intriguing result shown in Figure 12 is that even before reionization, a number of our halos exhibit significantly suppressed baryon fractions.  A similar result was seen by \citet{Simpson:2013} for high-resolution dwarfs and for \citet{Qin:2017} for large samples of dark matter halos. This is not a result of stellar feedback: the suppression is present even before star formation begins, and halo m10a, which forms no stars at any time, shows a significant (in fact the largest) reduction.  Our analysis suggests that the missing gas has been puffed out by shock heating -- many of these halos are experiencing rapid assembly (see \citealt{Wechsler:2002}).  They all reach the universal baryon fraction at a physical distance of 6 kpc from the halo center, which is significantly larger than the typical virial radii at these early times but is relatively small in terms of the distance that shock-heated gas can travel, even at early times (we note that the halos at early times have virial masses of $10^7-10^8\,\msun$, corresponding to virial temperatures of $\approx 4000-17000$ K and virial velocities of $\approx 10-20\,\kms$). In dwarf simulations performed in the absence of any ionizing background, we have found that the baryon fractions settle to much higher levels after the period of rapid merging ends (Elbert et al., in preparation), so this is distinct from the dominant late-time effect of reionzation. Future work will focus on the effects of reionization history on gas content and star formation histories in these dwarfs.

It is also interesting to note that while all of our simulated galaxies have baryon fractions that are significantly suppressed relative to the cosmic value of $\Omega_{\rm b}/\Omega_{\rm m}$, the final (virial) masses of the halos are essentially unaffected beyond baryon loss (i.e., $\mvir^{\rm DMO}/\mvir \approx 1$ after correcting the DMO virial mass for the cosmic baryon fraction; see columns 5 + 8 of Table~\ref{table:params}). This differs from some results in the literature. Both the Illustris \citep{Vogelsberger:2014b} and EAGLE \citep{Schaye:2015} simulations, which are hydrodynamic simulations with homogeneous mass resolution that cover large volumes at significantly lower resolution (dark matter particle masses of $\sim 10^6\,\msun$) than our zoom-ins, find that halos in the baryonic versions of their simulations have virial masses that are suppressed by an additional 10-15\% beyond the correction for $f_{\rm b}$ for low-mass halos (\citealt{Schaller:2015,Vogelsberger:2014a}, though Vogelsberger et al. find an average baryon fraction that is close to cosmic at the $10^{10}\,\msun$ scale). Sawala et al. also find a similar reduction in virial mass in simulations from both the GIMIC and APOSTLE projects \citep{Sawala:2013,Sawala:2016}. \citet{Munshi:2013} also see a larger reduction in halo mass in their baryonic zoom-in runs (see, e.g., their figure~5). The origin of this difference is not clear at present.

\subsection{Central densities of $\mstar \sim 10^6\,\msun$ dwarfs}
\label{subsec:previous}

The results presented in this paper solidify an emerging picture in which $\mvir(z=0) \approx 10^{10}\,\msun$, corresponding to $\mstar \sim 10^6\,\msun$, is a transition mass in \lcdm. More massive halos form more stars, with accompanying energy input from stellar feedback that converts dark matter cusps into cores. Lower-mass halos have substantially lower stellar masses, resulting in feedback input that is insufficient to modify CDM cusps \citep{Governato:2012,DiCintio:2014,Madau:2014,Chan:2015,Onorbe:2015,Tollet:2016}. The existence of a transition mass is likely related to the steep dependence of $\mstar$ on $\mvir$ in \lcdm\ simulations and abundance matching models: halos with only marginally different virial masses can vary by orders of magnitude in stellar content, meaning the ratio of stellar feedback energy to gravitational binding energy changes rapidly over a narrow range in halo masses. Our study marks a significant expansion in the exploration of halos that lie at the boundary of the cusp-core transition. We find that galaxies forming such halos can span nearly two decades in $\mstar$ (with one system remaining completely starless), yet this wide range of $\mstar(z=0)$ is somewhat deterministic: the amplitude of the central dark matter density in DMO versions of the simulations is an excellent predictor of the rank order of $\mstar(z=0)$. Equivalently, $\vmax$, concentration, and halo formation time all serve as proxies for stellar mass.

While our results on core formation are consistent with many recent cosmological simulations, they differ notably from \citet{Read:2016a,Read:2016b} and \citet{Sawala:2016}: Read et al. find cores at \textit{all} masses, while Sawala et al. do not find cores at \textit{any} mass. Read et al.'s simulations are non-cosmological, which requires the simulators to make a number of choices and assumptions about the initial conditions as well as the input physics. On the other hand, they are extremely high resolution, comparable to our ultra high-res Z14 simulations (for which we have only presented DMO results in this work;  Appendix~\ref{sec:appendixa}). Read et al. explore somewhat lower-mass halos ($\mvir \sim 10^8-10^9\,\msun$), yet the stellar masses line up well with the range simulated here: $\mstar=6.2\times 10^{5}\,\msun$ for $\mvir \approx 5\times 10^{8}\,\msun$ and $\mstar=3.6\times 10^{6}\,\msun$ for $\mvir \approx 10^{9}\,\msun$. Read et al. therefore find a very different $\mstar-\mhalo$ relation than we do at these masses. The absence of a UV background and no cosmological halo growth in the Read et al. simulation are likely to be the two most important sources of the differences seen relative to our simulations. 

As intimated in the Introduction, the differences we (and some other authors) find relative to Sawala et al. almost certainly have their roots in the treatment of star formation and feedback, as the background \lcdm\ cosmologies differ negligibly. The higher adopted value of $n_{\rm sf}$ and explicit treatments of energy injection from stellar evolution in the \fire-2 code are substantively different from the EAGLE/APOSTLE treatments, as are the inclusion of self-shielding of dense gas against the background UV field and the absence of an artificial temperature floor in our simulations. We believe that the modeling of these processes in the \fire-2 code is more realistic and is also well-converged numerically (\hopkins). Nevertheless, all cosmological simulations of galaxy formation are far from treating star formation in an ab initio manner; it is therefore crucial to understand which approximations are actually valid on galaxy-scale simulations. From the differences that Sawala et al. and Read et al. find, however, it is clear that observationally determining the presence or absence of cusps in galaxies with $\mstar \approx 10^{5}-10^{7}\,\msun$ is essential for understanding whether star formation feedback resolves various small-scale problems in \lcdm.

\cite{Onorbe:2015} hypothesized that core formation in $\mstar \sim 10^6\,\msun$ dwarfs is linked to late-time star formation, and \citet{Chan:2015} demonstrated that cores in similarly massive galaxies required repeated episodes of star formation feedback after the central gravitational potential stops growing (see also \citealt{Pontzen:2012}). In the halos that form enough stars to create dark matter cores, we do indeed see a correlation between the duration of star formation (as measured by the time when 50\% of star formation occurred) and the core density: halo m10i, which forms all of its stars in an early burst, does not see as much central density reduction as halo m10h, which forms the same amount of stars but over a much more extended period. This also tentatively supports the connection between late-time star formation and substantial core formation (subject, of course, to the overall mass in stars formed). This important question will best be answered by even larger samples of dwarfs at somewhat higher $z=0$ virial masses.

\section{Summary and Future Work}
\label{sec:summary}
We have simulated a suite of high-resolution, isolated dwarf galaxies, all having $\mvir(z=0)\approx 10^{10}\,\msun$, with the \gizmo\ code and the \fire-2 galaxy formation model. This is a mass scale that is of particular interest, both in terms of dwarf galaxies' susceptibility to UV background feedback and their ability to modify central dark matter cusps through star formation feedback. 
Our main results are as follows:
\begin{itemize}
\item Our halos, all chosen to have the same virial mass at $z=0$, have a variety of assembly histories. The assembly of the dark matter mass is highly correlated with the final stellar mass, especially when phrased in terms of $\vmax(z)$ (Figures~\ref{fig:mvir_vs_t} and \ref{fig:vmax_vs_t}).
\item A particularly good correlation to final stellar mass is found in the central density with the dark-matter-only version of each halo, with denser halos (in the DMO runs) forming more stars. At fixed halo mass, central density correlates strongly with $\vmax$, concentration, and formation time, meaning that we expect to see correlations with $\mstar$ and each of these properties at fixed halo mass. Central density in DMO simulations may therefore serve as a "second parameter" in setting stellar masses at fixed dark matter halo mass.
\item Our simulated galaxies have a variety of star formation histories, from solely high-redshift star formation in one case to late-time dominance in others. The star formation histories determined from the $z=0$ galaxies reproduce the diversity observed in Local Group dwarf galaxies.
\item Two galaxies in our sample self-quench and have no star formation (or cold gas) at $z=0$. These simulated galaxies are just below the mass scale at which \citet{Geha:2012} have observed a nearly complete absence of self-quenched galaxies, perhaps indicating that a population of quenched, low-mass dwarfs is waiting to be discovered (and potentially, that galaxies such as Cetus and Tucana self-quenched). This result is also consistent with the existence of the low-mass, quenched galaxies KKR 25 and KKs 3.
\item One of our halos fails to form any stars whatsoever. This halo is the latest-forming in our sample -- it has a very recent major merger -- and it lies below the characteristic UV suppression mass at all times.
\item We find a strong connection between total stellar mass at $z=0$ and the presence or absence of reduced central density: the galaxies that form more than $\approx 2\times 10^6\,\msun$ in stars all have reduced central dark matter densities, while those that fall below this stellar mass do not (Figures~\ref{fig:density_trippro} and \ref{fig:rho_dmo_vs_mstar}). This confirms the importance of $\mvir \approx 10^{10}\,\msun$ and $\mstar \approx 2 \times 10^{6}\,\msun$ for understanding whether the origins of cores lie in star formation feedback or dark matter physics beyond CDM.
\end{itemize}

The results presented in this paper cover only a subset of the interesting science related to the simulation suite we have introduced. It will also be important to explore halos slightly above and below the $\mvir \sim 10^{10}\,\msun$ transition mass studied here. On the more massive side, efficient density core creation should be common-place or ubiquitous; further tests of the correlation between core properties and dark matter assembly will provide insight into the core-cusp problem and the related issue of rotation curve diversity \citep{Oman:2015}. Slightly lower-mass halos should host galaxies with $\mstar \la 10^5\,\msun$, which can only be seen in the Local Group at present. Observations of these galaxies have revealed that they contain exclusively ancient stellar populations (e.g., \citealt{Brown:2012,Weisz:2014a}), which is often interpreted as a sign that reionization feedback controls the SFHs of these galaxies. Understanding the interplay between stellar and UV background feedback in such galaxies -- and confirming that such feedback is incapable of creating cores in systems at the low-mass edge of galaxy formation -- will lay the groundwork for direct tests of the \lcdm\ model.

\section*{Acknowledgments}
MBK and AF acknowledge support from the National Science Foundation (grant AST-1517226). MBK was also partially supported by NASA through HST theory grants (programs AR-12836, AR-13888, AR-13896, and AR-14282) awarded by the Space Telescope Science Institute (STScI), which is operated by the Association of Universities for Research in Astronomy (AURA), Inc., under NASA contract NAS5-26555. JSB and ODE were supported by  NSF AST-1518291, HST-AR-14282, and HST-AR-13888. Support for PFH was provided by an Alfred P. Sloan Research Fellowship, NASA ATP Grant NNX14AH35G, and NSF Collaborative Research Grant \#1411920 and CAREER grant \#1455342. AW was supported by a Caltech-Carnegie Fellowship, in part through the Moore Center for Theoretical Cosmology and Physics at Caltech, and by NASA through grant HST-GO-14734 from STScI. DK was supported by NSF grant AST-1412153, a Cottrell Scholar award, and funds from the University of California, San Diego. CAFG was supported by NSF through grants AST-1412836 and AST-1517491, by NASA through grant NNX15AB22G, and by STScI through grant HST-AR-14293.001-A. 

This work used computational resources of the University of Maryland, The University of Texas at Austin and the Texas Advanced Computing Center (TACC; \url{http://www.tacc.utexas.edu}), the NASA Advanced Supercomputing (NAS) Division and the NASA Center for Climate Simulation (NCCS) through allocations SMD-15-5902, SMD-15-5904, SMD-16-7043, and SMD-16-6991, and the Extreme Science and Engineering Discovery Environment (XSEDE, via allocations TG-AST110035, TG-AST130039, and TG-AST140080), which is supported by National Science Foundation grant number OCI-1053575. 

\bibliography{draft}

\appendix
\section{Resolution and Convergence}
\label{sec:appendixa}
\begin{figure*}
\includegraphics[width=1.0\textwidth]{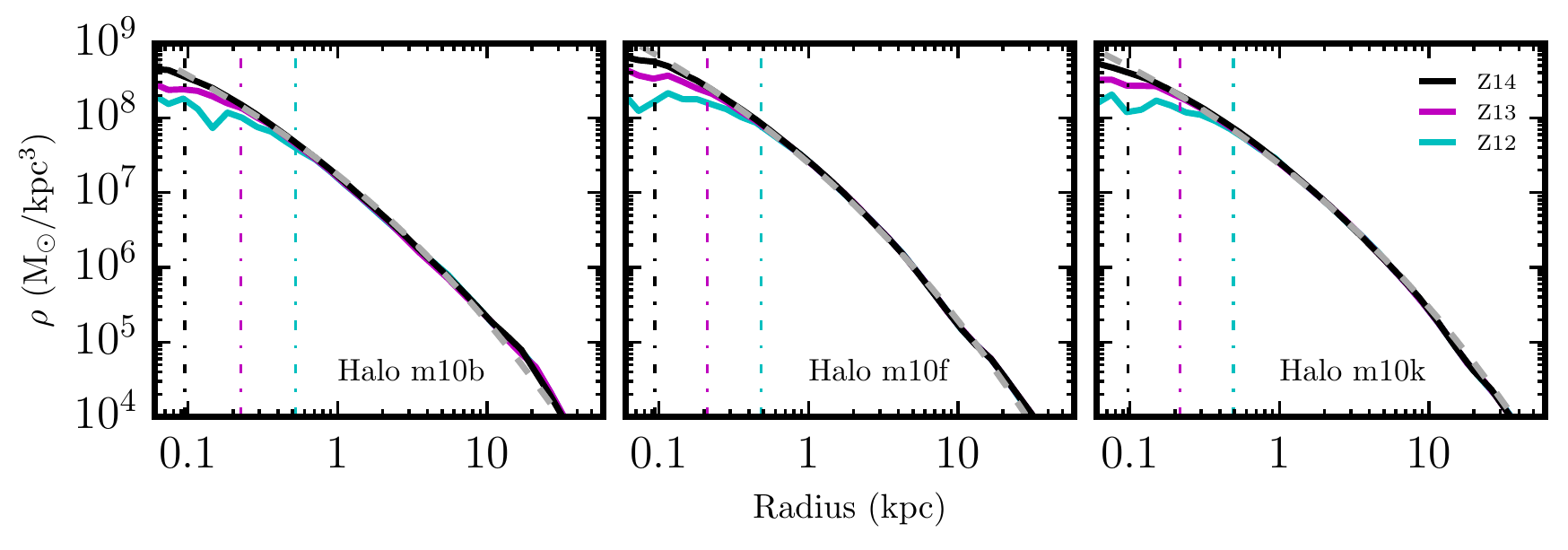}
  \caption{Numerical convergence for DMO simulations. Each panel shows the density profile for DMO runs of an individual halo at three resolutions: Z12 (cyan), Z13 (magenta), and Z14 (black). The Power radius for each run is marked by a vertical dotted line of the corresponding color and provides a relatively conservative approximation for where each density profile deviates from its higher resolution counterpart (i.e., density profiles are essentially perfectly converged for $r \ge \rpower$ and are converged to better than $\sim 20\%$ in density for $r \ga 0.5 \,\rpower$). At our fiducial resolution (Z13),  $\rpower$ is $\approx 200\,{\rm pc}$ for the DMO simulations. The gray dashed line in each panel shows the best-fitting Einasto profile (with $\alpha$ fixed to 0.17) for the Z14 run; Einasto profiles provide a good description of the density profiles in our simulations.
}
  \label{fig:dmo_converge}
\end{figure*}
\begin{figure*}
\includegraphics[width=1.0\textwidth]{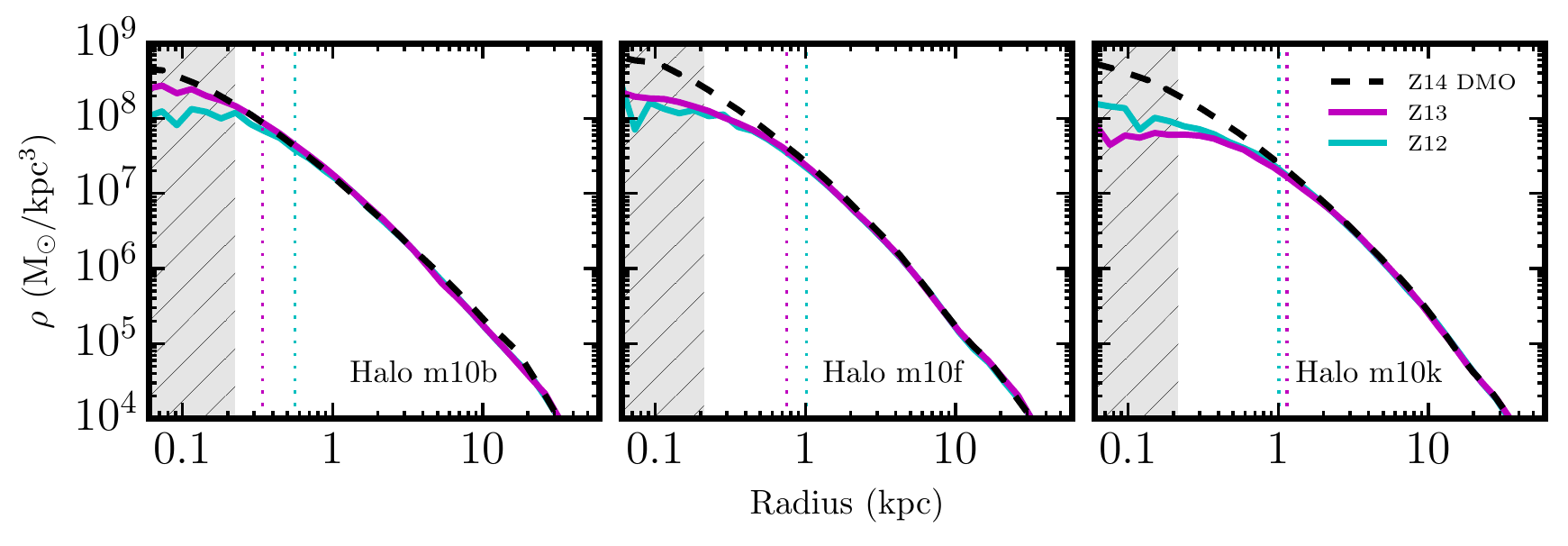}
  \caption{Density profile convergence between our fiducial resolution (Z13, magenta curves) and lower resolution (Z12, cyan curves) counterparts in the full physics simulations for the same three simulations as in Figure~\ref{fig:density_trippro}. The ultra-high resolution DMO density curves are also plotted as black curves for comparison. Vertical dotted lines mark the half-mass radius, while the gray hatched region shows where numerical relaxation may affect the Z13 results according to the Power et al. criterion. There is excellent convergence in the density profiles of the hydrodynamical runs across resolution levels. For reference, the fractional change in $z=0$ stellar mass from Z12 to Z13 is, from left to right, 1.1, 1.3 and 2.3 (i.e., lower-resolution simulations form somewhat fewer stars).}
  \label{fig:density_converge_hydro}
\end{figure*}
\begin{figure}
\includegraphics[width=\columnwidth]{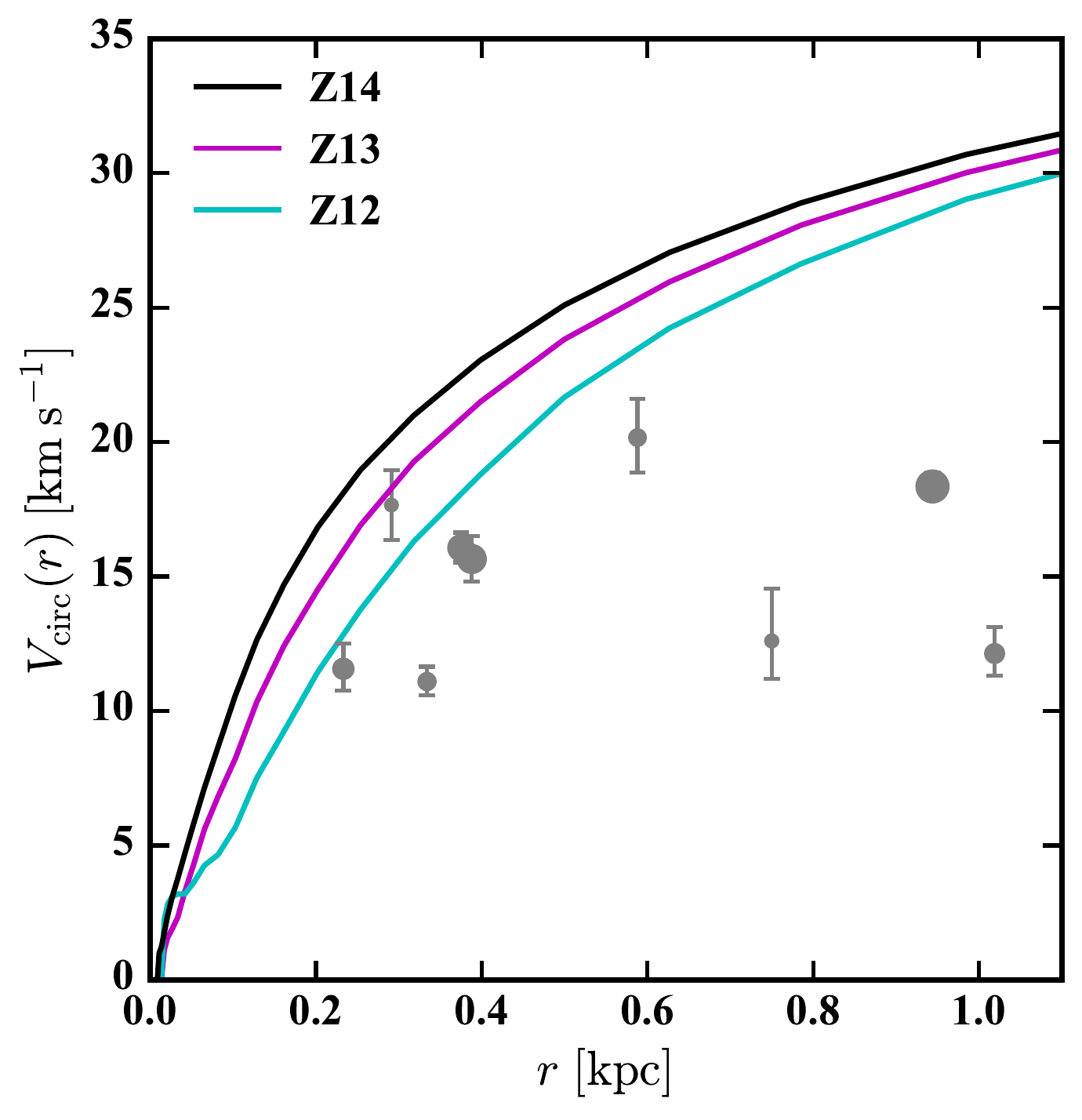}
  \caption{Importance of resolution in DMO simulations. The simulated circular velocity profile of one halo simulated at three resolutions (low: Z12, cyan; fiducial: Z13, magenta; ultra-high: Z14, black) is plotted as a function of radius. Symbols with error bars show the measured circular velocities of the nine bright MW dwarf spheroidal galaxies, which are relevant for the "too big to fail" problem. Differences of $5\,\kms$ from low to high resolution are present in the inner 200-400 pc. We note that (1) the density profiles are all converged within $\sim 400$ pc (Fig.~\ref{fig:dmo_converge}), yet the circular velocities differ substantially (owing to the cumulative nature of $\vcirc$); and (2) even our "low resolution" simulation uses dark matter particle masses of $\approx 2.5\times 10^{4}$, which is smaller (better resolved) than any published cosmological simulation of the formation of the Milky Way and its satellites.}
  \label{fig:dmo_vcirc}
\end{figure}
Fig.~\ref{fig:dmo_converge} examines convergence of density profiles in 3 DMO simulations at three resolution levels (seperated by a factor of 64 in mass between our lowest resolution -- Z12, in cyan -- and our highest resolution, Z14, which is plotted in black). \citet{Power:2003} proposed that an estimate of numerical convergence radius for density profiles in dark matter simulations is the radius where the two-body relaxation time exceeds 60\% of the current age of the Universe (corresponding to the radius enclosing $\sim2500$ particles); Fig.~\ref{fig:dmo_converge} demonstrates that this Power criterion provides a conservative measure of numerical convergence. We refer to this ``Power radius'' (calculated just from dark matter particles) as our reference ``convergence radius'' throughout (and note that $\sim20\%$ convergence in density can be obtained at radii enclosing just $\sim200$ particles). The Power radius for each simulation is marked with a dotted line, with color matching the corresponding density profile, in the figure.  Fig.~\ref{fig:dmo_converge} also shows the best-fitting Einasto profiles (with $\alpha$ fixed to 0.17) for the Z14 simulations (gray dashed lines). In each case, the Einasto profile provides an excellent fit for all converged radii (small fluctuations at large radii are due to substructure).

Figure~\ref{fig:density_converge_hydro} shows density profiles from hydrodynamic runs of the same three halos plotted in Figure~\ref{fig:dmo_converge} at fiducial (Z13; magenta) and low (Z12; cyan) resolution. We reiterate that no parameters related to star formation or feedback are changed between the two different hydrodynamic resolution levels, making for a clean comparison. We also plot the ultra-high resolution (Z14) DMO density profile in each case (black dashed curve). The gray hatched region shows where the Power criterion indicates results at our fiducial resolution may not have converged in DMO runs. In each case, the density profiles agree well between the two resolutions for all converged radii. The 3D stellar half-mass radii, marked by vertical dotted lines, agree well for the most part, too (the smallest galaxy, m10b, is approximately 50\% larger in the lower resolution simulation). Across our sample, we generally find that stellar masses increase by a median of 40\% when moving from low (Z12) to fiducial (Z13) resolution, with only one galaxy having a lower stellar mass at higher resolution. Given the complex physical phenomena at work and the change by a factor of 8 in particle masses across resolution levels, we find this agreement to be encouraging. Production runs at Z14 with identical implementations of hydrodynamics and galaxy formation physics are in the planning stage and will be presented in future papers. These runs will provide an even stronger test of numerical convergence.

Figs.~\ref{fig:dmo_converge} and \ref{fig:density_converge_hydro} explore convergence in density profiles. Convergence of circular velocity profiles $\vcirc(r)$ is slower -- that is, convergence in $\vcirc$ is generally less good than in $\rho$ at fixed radius -- because $\vcirc$ a cumulative quantity. The convergence in $\vcirc$ of our DMO simulations is shown in figure~\ref{fig:dmo_vcirc} and emphasizes this point: 
the highest and lowest resolution runs here differ by as much as $\approx 5\,{\rm km\,s^{-1}}$ at $\sim200-400\,$pc even though the density profiles are almost perfectly converged outside this radius. 
This slower convergence is important: if we compare the observed central $\vcirc$ (at the half-light radii) of the 9 brightest dSph satellites of the MW (from \citealt{Boylan-Kolchin:2012}), we see that the low-resolution run is consistent with at least two satellites while the highest-resolution run is inconsistent (denser) than all satellites (note that our standard correction for $f_{\rm b}$ has been applied to all three runs). We emphasize that our lowest-resolution run here is actually higher resolution than any published cosmological simulation of the MW and its satellites, highlighting the difficulty of simulating both the MW and its satellite system in hydrodynamic run. However, we also note that the difference shown is larger than that in some of our other DMO simulations. Furthermore, the effect shown here can be subdominant to baryonic effects on mass profiles in hydrodynamical simulations with core formation, leading in many cases to more rapid convergence (less resolution dependence). These points are explored in further detail in \hopkins.

\section{Galaxy stellar mass definition}
\label{sec:appendixb}
\begin{figure}
\includegraphics[width=\columnwidth]{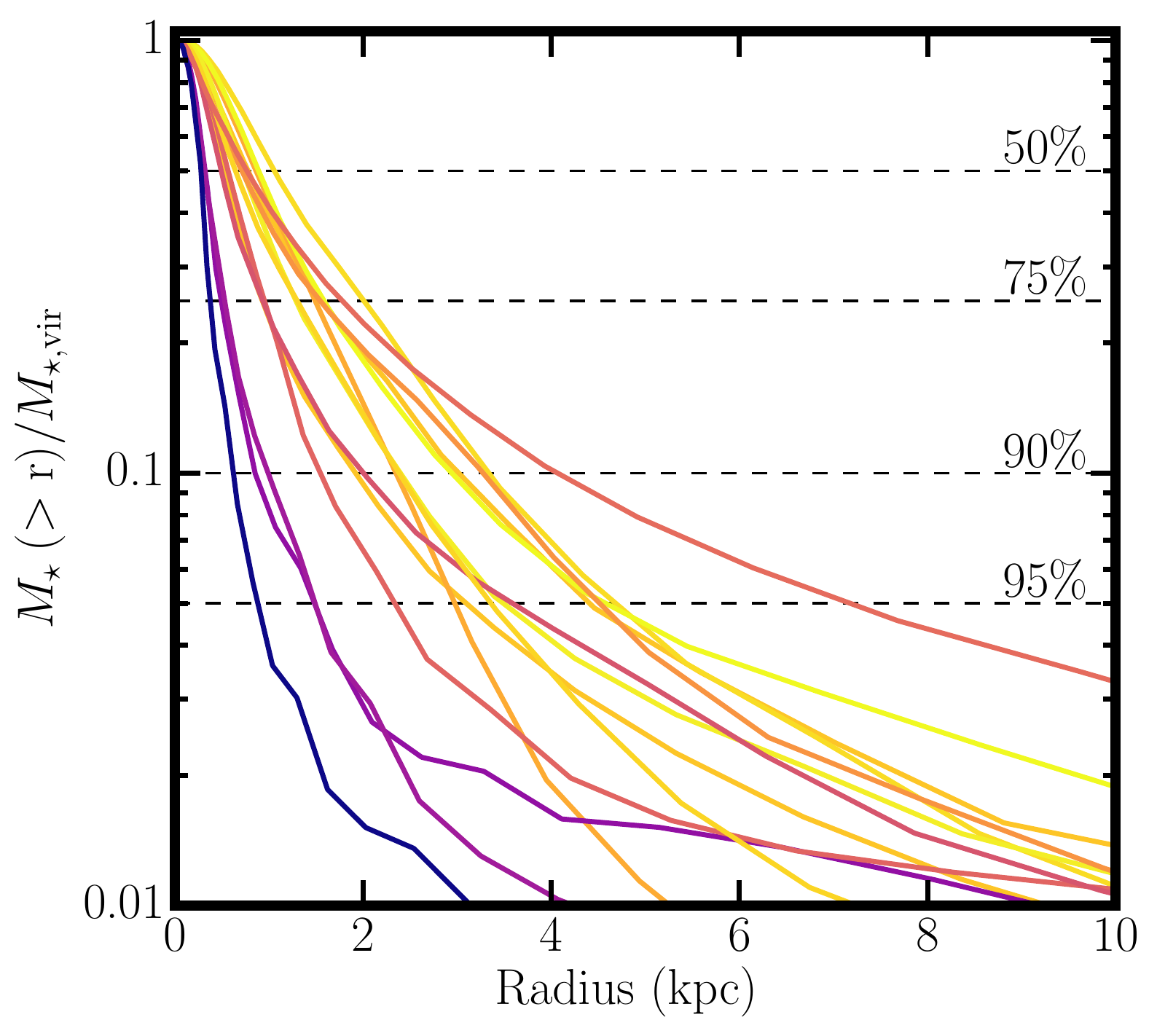}
    \vspace{-0.5cm}
  \caption{Stellar mass profiles of our simulated galaxies, plotted as the fraction of stars within $\rvir$ external to radius $r$. Line coloring indicates galaxy mass, with the same scale as Fig.~\ref{fig:mvir_vs_t}, and horizontal lines show fixed fractions of the total stellar mass (e.g., the 95\% line indicates the radius $r$ where 95\% of the halo's stellar mass is contained within $r$). In all cases, 95\% (90\%) of the stellar mass is contained within 7 (4) kpc. We define the stellar mass of a galaxy to be the mass in stars contained within $0.1\,\rvir \approx 6\,{\rm kpc}$. Any choice between $\sim 3\,{\rm kpc}$ and $\rvir$ will result in very similar measurements of stellar properties of our galaxies.}
  \label{fig:mstar_vs_r}
\end{figure}
There is no unique way to define the stellar mass of a simulated galaxy. Common choices include taking all stellar mass within the virial radius -- which is reasonable for dwarfs, as satellites contribute very little stellar mass -- or the stellar mass within a fixed radial aperture of $\sim 2-5\,{\rm kpc}$. Figure~\ref{fig:mstar_vs_r} demonstrates the ambiguity in our simulations: the fraction of stellar mass within $\rvir$ external to a radius $r$ is plotted as a function of $r$. The lines are colored according to total stellar mass. The galaxies have a wide range of profiles at large radii, from relatively sharply truncated to very extended. Nevertheless, each halo contains at least 90\% of its stars within 4 kpc of its center, meaning the extended, low surface brightness wings do not affect stellar mass (or stellar half-mass radius) measurements appreciably. For concreteness, we define stellar mass to be the mass contained within $0.1\,\rvir$ ($\approx 6\,{\rm kpc}$ for the halo mass scale studied here). This comprises between 92 and 100\% of the total stellar mass within $\rvir$ for all halos.

\bsp	
\label{lastpage}
\end{document}